\documentclass{ieeeaccess}
\usepackage{cite}
\usepackage{amsmath,amssymb,amsfonts}
\usepackage{algorithmic}
\usepackage{graphicx}
\usepackage{textcomp}
\usepackage{multirow}
\usepackage{pythonhighlight}
\usepackage{makecell}
\def\BibTeX{{\rm B\kern-.05em{\sc i\kern-.025em b}\kern-.08em
    T\kern-.1667em\lower.7ex\hbox{E}\kern-.125emX}}
\begin{document}
\history{Date of publication xxxx 00, 0000, date of current version xxxx 00, 0000.}
\doi{10.1109/ACCESS.2021.3050757}

\title{An Optimal Energy-Saving Home Energy Management Supporting User Comfort and Electricity Selling with Different Prices}
\author{\uppercase{HUY TRUONG DINH}, \IEEEmembership{Student Member, IEEE},
\uppercase{and DAEHEE KIM},
\IEEEmembership{Member, IEEE}}
\address{Department of Future Convergence Technology, Soonchunhyang University, Asan 31538, South Korea}

\tfootnote{This work was supported by the Korea Institute of Energy Technology Evaluation and Planning (KETEP) and the Ministry of Trade, Industry \& Energy (MOTIE) of the Republic of Korea (No. 20184030202130) and this work was supported by the Soonchunhyang University Research Fund.}

\markboth
{Author \headeretal: Preparation of Papers for IEEE TRANSACTIONS and JOURNALS}
{Author \headeretal: Preparation of Papers for IEEE TRANSACTIONS and JOURNALS}

\corresp{Corresponding author: Daehee Kim(e-mail: daeheekim@sch.ac.kr)}

\begin{abstract}
In this study, we investigate the operation of an optimal home energy management system (HEMS) with integrated renewable energy system (RES) and energy storage system (ESS) supporting electricity selling functions. A multi-objective mixed integer nonlinear programming model, including RES, ESS, home appliances and the main grid, is proposed to optimize different and conflicting objectives which are energy cost, user comfort and PAR. The effect of different selling prices on the objectives is also considered in detail. We further develop a formula for the lower bound of energy cost to help residents or engineers quickly choose best parameters of RES and ESS for their homes during the installation process. The performance of our system is verified through extensive simulations under three different scenarios of normal, economic, and smart with different selling prices using real data, and simulation results are compared in terms of daily energy cost, PAR, user's convenience and consecutive waiting time to use appliances. Numerical results clearly show that the economic scenario achieves $51.6\%$ reduction of daily energy cost compared to the normal scenario while sacrificing the user's convenience, PAR, and consecutive waiting time by $49\%$, $132\%$, and 1 hour, respectively. On the other hand, the smart scenario shows only slight degradation of user's convenience and PAR by $2\%$ and $18\%$, respectively while achieving $46.4\%$ reduction of daily energy cost and the same level of consecutive waiting time. Furthermore, our simulation results show that a decrease of selling prices has tiny impacts on PAR and user comfort even though the daily energy cost increases.
\end{abstract}


\titlepgskip=-15pt

\maketitle

\section{Introduction}
\label{sec:introduction}
\PARstart{I}{n} the future, the smart grid (SG) will not only be a main component of electricity delivery between suppliers, prosumers and consumers but also play a key role in reducing energy consumption \cite{dileep2020survey}. By using advanced metering infrastructure (AMI) of the SG, residents can utilize external information sent by the utilities to improve their energy usage. On the other side, the SG also helps the utilities distribute power in more effective way and reduce the system's peak-to-average ratio (PAR), a main contributing factor to an electricity crisis \cite{liu2014peak}. With the amazing development of renewable energy and energy storage technologies, more and more houses are able to set up home energy management systems (HEMSs) with integrated renewable energy systems (RESs) and energy storage systems (ESSs) to reduce their energy cost, PAR, and dependency on the main grid (MG) in an efficient and reliable manner. In the SG, participating in the electricity market is another efficient way to decrease energy costs and maximally utilize renewable energy from residential zones. Hence, the main role of HEMS is not only to control and manage all electrical devices, but also to fully support selling operations by the residents, fulfilling their various requirements. Due to the tremendous effects of HEMS on energy consumption of houses, there are numerous studies on the different problems HEMS may encounter. A large number of optimization models and scheduling schemes have also been proposed. For example, in \cite{althaher2015automated}, Sereen Althaher \textit{et al.} proposed an optimization-based automated demand response (ADR) to be implemented in HEMS. Their ADR aimed to reduce the consumer's electricity bill below a certain level, whilst increasing their comfort. To achieve these objectives, a mixed integer nonlinear programming (MINLP) model for optimizing the cost of energy with thermal and device constraints was built. However, in their study, RES and ESS were not considered, and thus selling operations were not mentioned.

The authors in \cite{nguyen2013joint} presented a joint optimization of electric vehicles (EV) and home energy scheduling. In their study, EV was exploited as dynamic energy storage. Their objectives  were to minimize energy cost while considering thermal constraints and the constraints of EV travel. Utilization of the MG and selling activities were mentioned to reduce their energy cost during high price slots. To solve their optimization problem, they employed the Model Predictive Control (MPC) technique. However, the set of appliances in the house was not considered. Thus, the detailed schedule of each appliance and selling operations in each time slot were not given consideration. Moreover, RES was not considered in this paper. 

In \cite{shakeri2017intelligent}, a HEMS integrated battery storage system and photo-voltaic system were presented. Their objectives were to minimize energy cost and satisfy the thermal and device constraints. In their study, an example set of appliances for a house was given. Instead of building an optimization model, they proposed a load management algorithm. Their algorithm only controlled battery and thermal appliances to adapt to operations of other appliances which were turned on or off by residents. In their work, selling was not mentioned.

In \cite{samadi2015load}, an energy management system with integrated RES and ESS for a group of homes was proposed to optimize energy cost and allow power trading. At each home, a set of appliances was given and a model was also built for minimizing energy cost and scheduling these appliances. In this paper, selling operations between the homes was supported using the Nash equilibrium from game theory. However, utilization of the MG and user comfort were not considered for each home in their model. 

Similar to \cite{samadi2015load}, in \cite{wang2018pareto}, multiple-home energy management system was proposed. In this system, three types of home were introduced, including a home with both an EV and a PV, a home with a single EV, and a home without EV and PV. Maximization of consumer satisfaction, minimization of energy cost, and minimization of PAR were considered as their objectives. To solve their multi-objective optimization problem, pareto tribe evolution with Nash equilibrium-based decision (PTE-Nash) was used. However, selling operation was not considered in their system.

In \cite{anvari2014optimal}, \cite{anvari2017efficient}, and \cite{anvari2017multi}, the authors built an optimization model for HEMS with integrated RES and ESS to optimize energy cost and user comfort. A multi-objective MINLP optimization model was developed for considering both energy saving, thermal comfort and the user's convenience. A detailed schedule of appliances was given by solving this MINLP model. Although selling activities were mentioned, their HEMS failed to give a detailed schedule for selling operations in each time slot. Moreover, the effect of selling price on system performance and user comfort were not considered.

In \cite{ahmed2017real}, authors developed an optimal real time schedule controller for HEMS to minimize energy consumption during a week. In their study, a set of home appliances was given. A new binary backtracking search algorithm (BBSA) was introduced to solve their optimization problem and the performance of this new algorithm was compared with binary particle swarm optimization (binary PSO). Simulation results showed that the performance of the BBSA controller was better in terms of saving energy compared to the binary PSO controller. However, RES and ESS were not considered in their study.

In \cite{bouakkaz2020efficient}, an HEMS with integration of RES and battery energy storage system (BESS) was proposed to reduce energy cost in a home. A given set of home appliances included six shiftable appliances and six fixed appliances. In their study, a mathematical model of the system was built. In this model, BESS energy was supposed to be not free and its price depends on the state of charge (SOC). This price allowed the optimal use of battery energy, thus enhancing the battery lifetime. To solve the optimization problem, PSO was used. Three scenarios were introduced: base scenario without scheduling, optimization with scheduling appliances (OSA), and optimization with scheduling appliances and batteries (OSAB). Simulation results proved that OSAB was more effective in reducing the energy cost and BESS energy loss than other scenarios. However, in this study, user comfort and electricity selling were not considered.

In \cite{ahmed2016artificial}, a home energy management controller was proposed using artificial neural network (ANN) to decrease the energy consumption for home devices at specific times. A example set of home devices included air conditioner, electric water heater, washing machine and refrigerator. Their ANN was trained to predict the optimal ON/OFF status of the home devices. Results showed that the proposed ANN based controller can reduce the energy consumption and maintain the total power consumption below a threshold without affecting resident lifestyles. However, RES and ESS were not considered in their study. 

In \cite{hemmati2017stochastic}, an HEMS with integrated BESS and PV system was presented. In this study, BESS and PV system were utilized to reduce electricity bill in smart home. The proposed planning for determining optimal strategy was expressed as a stochastic MINLP which was solved by advanced-adaptive PSO (AAPSO). In their HEMS, selling and buying operation from network were supported. However, selling price equaled to buying price and both prices were the same as main grid price. Hence, the effect of different selling prices was not considered in this study. Moreover, scheduling for appliances was not considered.

In our previous work \cite{dinh2020home}, we presented a HEMS to optimize energy cost and PAR. To achieve these objectives, optimization formulas were built and solved using a PSO algorithm. A detailed schedule of utilization of the MG and selling operations were given for each time slot but user comfort was not considered. 

Beside \cite{hemmati2017stochastic} and \cite{dinh2020home}, there are many other studies that use meta-heuristic algorithms for scheduling appliances such as genetic algorithm (GA), wind driven optimization (WDO), harmony search algorithm (HSA), and so on \cite{shaikh2016intelligent}, \cite{ahmad2017optimized}, \cite{hussain2018efficient}, \cite{rahim2016exploiting}, \cite{javaid2018day}, \cite{rahim2018energy}, \cite{awais2018towards}, \cite{samuel2018efficient}. Although energy cost and various kinds of user comfort were considered and optimized in these studies, all of them failed to consider the selling operation in their optimization problems. 

Two main techniques widely used to solve optimization problem of a HEMS in a home are meta-heuristic algorithms and mathematical optimization algorithms. Each kind of techniques has its own advantages and disadvantages, and depends on the environment in which a HEMS is built. However, in meta-heuristic algorithms, optimal values need to be investigated for many parameters such as swarm size, number of iterations, and so on. Moreover, in our previous work, meta-heuristic algorithms take a lot of time to solve our optimization problem and output optimal results. Hence, in this work, mathematical optimization algorithms are selected to solve our problem.

\begin{table*}[]
\caption{A comparison of HEMS: State of the art}
\begin{center}
\begin{tabular}{|c|c|c|c|c|c|c|c|c|c|c|c|}
\hline
\multirow{3}{*}{\textbf{Work}}&\multirow{3}{*}{\makecell{\textbf{Algorithm} \\ \textbf{(Technique)}}} & \multicolumn{2}{c}{\textbf{Integration}} &\multicolumn{4}{|c|}{\textbf{Objective}}&\multirow{3}{*}{\makecell{\textbf{Consecutive} \\ \textbf{Constraints}}} & \multirow{3}{*}{\makecell{\textbf{Utilizing} \\ \textbf{MG}}} & \textbf{Different} &\textbf{Lower Bound}\\
\cline{3-8} 
& &\multirow{2}{*}{\textit{ESS}} &\multirow{2}{*}{\textit{RES}} &\textit{Energy} &\multirow{2}{*}{\textit{PAR}} &\textit{User} &\textit{Consecutive} & && \textbf{Selling} &\textbf{of}\\
& & & &\textit{Cost} &  &\textit{Convenience} &\textit{Waiting Time} & & &\textbf{Prices} & \textbf{Energy Cost}\\
\hline
\cite{althaher2015automated} & MINLP & & & \checkmark & & & & & & & \\
\hline
\cite{nguyen2013joint} & MPC & \checkmark & &\checkmark & & & & &\checkmark &  & \\
\hline
\cite{shakeri2017intelligent} & load control & \checkmark & \checkmark & \checkmark & & & & & & & \\
\hline
\cite{samadi2015load} & game theory & \checkmark & \checkmark & \checkmark & & & & & &\checkmark & \\
\hline
\cite{wang2018pareto} & PTE-Nash & \checkmark & \checkmark & \checkmark & \checkmark & & & & & & \\
\hline
\cite{anvari2014optimal} & MINLP & \checkmark &  & \checkmark &  &\checkmark & & \checkmark & \checkmark & & \\
\hline
\cite{anvari2017efficient} & MINLP & \checkmark & \checkmark & \checkmark &  &\checkmark & &  \checkmark & \checkmark & & \\
\hline
\cite{ahmed2017real} & BBSA & & & \checkmark & & & & & & & \\
\hline
\cite{bouakkaz2020efficient} & meta-heuristic & \checkmark & \checkmark & \checkmark & & & & & & & \\
\hline
\cite{ahmed2016artificial} & ANN & & & \checkmark & & \checkmark & & & & & \\
\hline
\cite{hemmati2017stochastic} & meta-heuristic & \checkmark & \checkmark & \checkmark & & & & & & & \\
\hline
\cite{shaikh2016intelligent} & meta-heuristic & & &\checkmark & & & & & & & \\
\hline
\cite{ahmad2017optimized} & meta-heuristic  &\checkmark &\checkmark & \checkmark & \checkmark & & & & & & \\
\hline
\cite{hussain2018efficient} & meta-heuristic & & & \checkmark & \checkmark & \checkmark  & & & & & \\
\hline
\cite{rahim2016exploiting} & meta-heuristic & \checkmark & \checkmark & \checkmark & \checkmark & \checkmark & & & & & \\
\hline
\cite{javaid2018day} & meta-heuristic & & & \checkmark & \checkmark & \checkmark & & & & & \\
\hline
\cite{rahim2018energy} & meta-heuristic & & & \checkmark & \checkmark &\checkmark & & & & & \\
\hline
\cite{awais2018towards} & meta-heuristic & & & \checkmark & \checkmark &\checkmark & & & & & \\
\hline
\cite{samuel2018efficient} & meta-heuristic & \checkmark & \checkmark & \checkmark & \checkmark & \checkmark & & & & & \\
\hline
Previous work & meta-heuristic & \checkmark & \checkmark & \checkmark & \checkmark & & & & \checkmark & \checkmark &\\
\hline
Our work & MINLP & \checkmark & \checkmark & \checkmark & \checkmark &\checkmark & \checkmark & \checkmark & \checkmark & \checkmark & \checkmark\\
\hline
\end{tabular}
\label{survey_table}
\end{center}
\end{table*}

Motivated by the above literature works, we propose a novel HEMS with integration of RES and ESS supporting user comfort and electricity selling. In this HEMS which is an extension of our previous work \cite{dinh2020home}, the effects of different selling prices on energy cost, user comfort and PAR are fully considered. In this study, a MINLP model, including different objectives: energy cost per day, user comfort, and PAR, is modeled. Two kinds of user comfort are considered in our model: the user's convenience and consecutive waiting time to use appliances. To the best of our knowledge, none of the previous studies considered an optimization model which takes daily energy cost, user comfort, and PAR into account and supports electricity selling. The brief comparison of research works on HEMS is listed in Table \ref{survey_table}. The main contributions of our works can be summarized as follows.

\begin{itemize}
\item  A multi-objective MINLP model that jointly optimizes four different objectives is built. This model aims to achieve a balance between minimizing energy cost and preservation of user comfort and PAR. A detailed schedule for the operation of appliances, utilization of the MG and selling operation in each time slot are given to achieve this optimal balance. The effect of selling price on system performance is considered. Specifically, simulation results show that a decrease of selling price impacts user comfort and PAR slightly.
\item A formula for the lower bound of energy cost is developed. Due to its simplicity, this formula can help residents or engineers quickly estimate the economic benefit achieved by using a HEMS with integrated specific RES and ESS. Thus, residents or engineers can easily choose the parameters of the RES and ESS that are the best fit for their homes. The numerical results show that the minimum value of our energy cost approaches very close to this lower bound.
\item We show the impacts of different weight coefficients, which control energy cost, user comfort, and PAR, on system operation as well as the economic benefits. The weight method for multi-objective optimization allows us more flexibility in setting trade-offs between energy cost, user comfort, and PAR.
\end{itemize}
The remainder of this paper is organized as follows. A brief description of the HEMS system is shown in Section II. A detailed problem formulation and optimization model are built in Section III. Section IV investigates the lower bound of energy cost. In Section V, the scenarios and simulation results are provided. Finally, Section VI outlines conclusions and potential future works.
\section{System Description}
\begin{figure}[!t]
\centering
\includegraphics[scale=0.4]{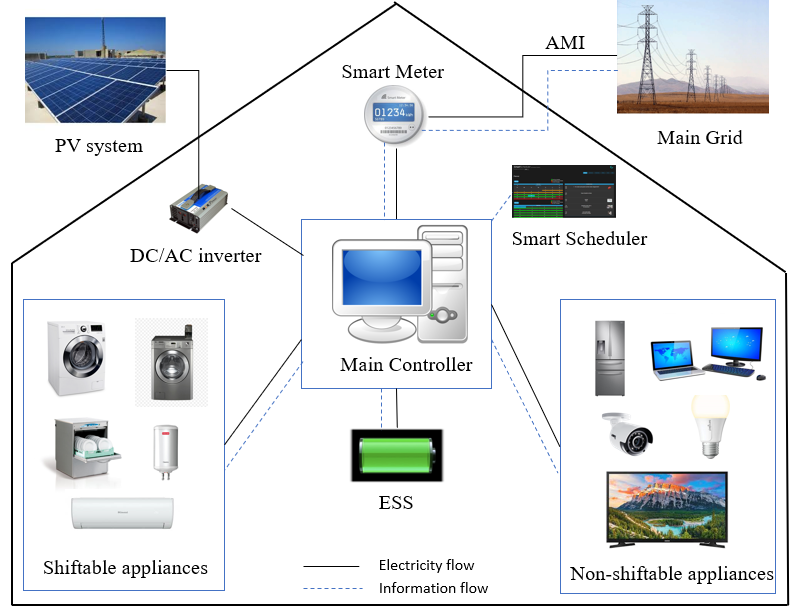}
\caption{HEMS Architecture \cite{dinh2020home}.}
\label{HEMS_architecture}
\end{figure}

\begin{figure}[!t]
\centering
\includegraphics[scale=0.9]{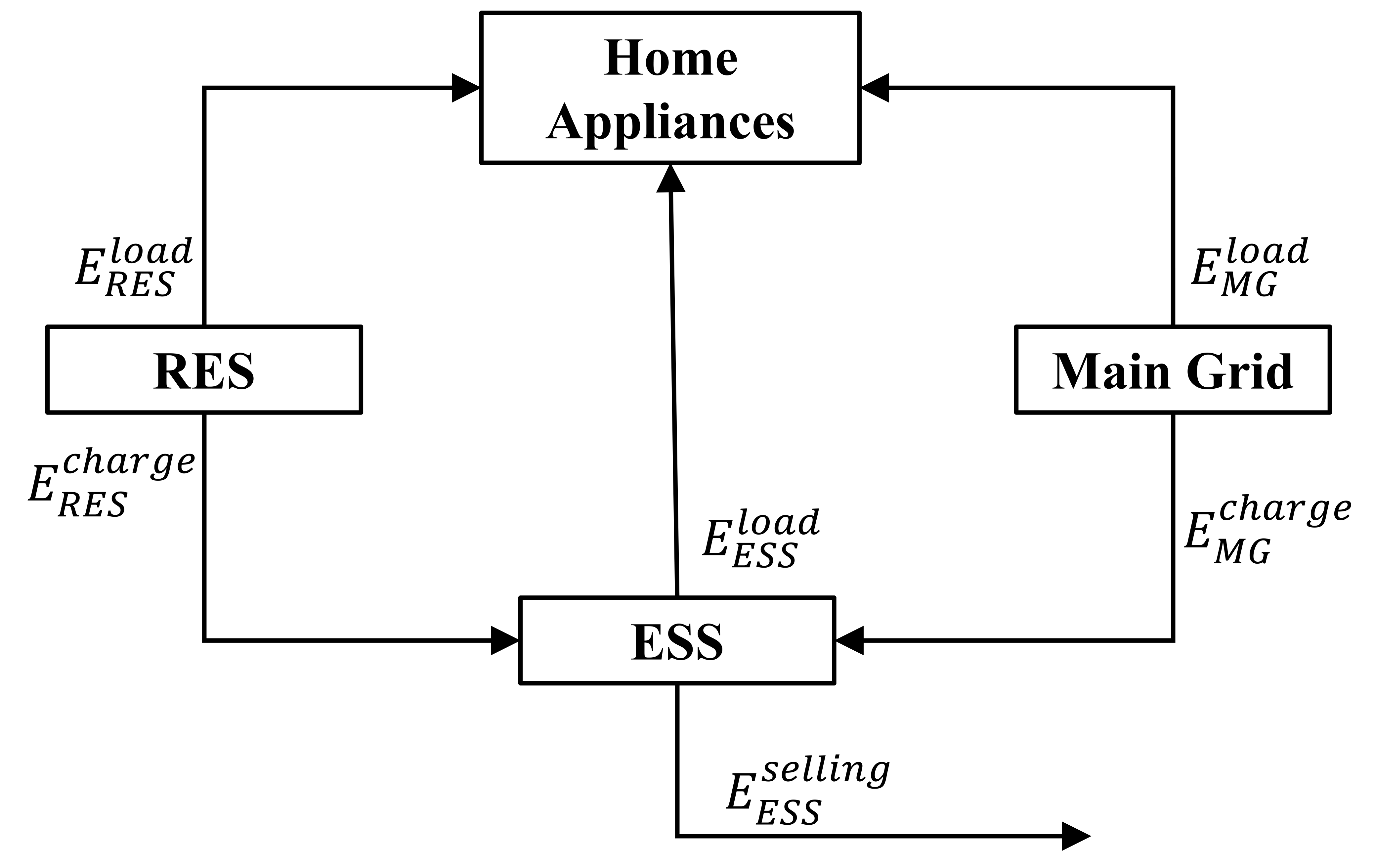}
\caption{Energy flows in our HEMS \cite{dinh2020home}.}
\label{electricity_flow}
\end{figure}

In this paper, as shown in Fig. \ref{HEMS_architecture}, a smart home with HEMS and a collection of shiftable and non-shiftable electrical appliances is studied. Generally, key components of a HEMS include AMI, a main controller (MC), ESS and RES. External information can be collected through an AMI. This includes pricing information, forecast temperature, and solar irradiance. The MC uses this useful information to control all electrical devices including the ESS and RES. The RES is used to decrease the dependency on the MG and reduce energy cost. In this paper, we assume that a PV system is set up as the RES. To be able to store surplus RES energy and utilize electricity from the MG at low price times, the ESS is needed for our system. Moreover, our HEMS supports residents in selling the electricity. Fig. \ref{electricity_flow} shows all energy flows in our smart home.
\section{Problem Formulation}
In this section, we build mathematical formulas for the RES, ESS, appliances and our objectives during a day time from 0 A.M. to 12 P.M. We divide a day into $T=24$ time slots and the duration of each time slot is $\Delta t=1h$. It is important to note that the following formulas are valid even when $\Delta t$ is smaller and $T$ is bigger.

\subsection{Renewable Energy Source}

In this work, our HEMS is equipped with a PV system as its RES. According to \cite{dinh2020home}, the output energy, $E_{RES}(t)$, from a PV system in kWh in any time slot $t$ ($1 \leq t \leq T$) can be measured as 

\begin{equation}
E_{RES}(t)= GHI(\tau) \cdot S \cdot \eta^{RES} \cdot \Delta t.
\label{RES_power}
\end{equation}
where $GHI$ is the global horizontal irradiation $(kW/m^2)$ at the location of the solar panels. $\tau$ is the real time in time slot $t$. $S$ is the total area $(m^2)$ of solar panels and $\eta^{RES}$ is the solar conversion efficiency of the PV system.

As shown in Fig. \ref{electricity_flow}, this energy would be used for home load and ESS charging. Thus, we have the following equation.

\begin{equation}
E_{RES}(t) = E_{RES}^{load}(t) + E_{RES}^{charge}(t)
\label{RES_elements_1}
\end{equation}
where $E_{RES}^{load}(t)$ is the energy quantity used for home load in time slot $t$. $E_{RES}^{charge}(t)$ is the energy quantity used to charge the ESS in time slot $t$. In real life, because every ESS is only able to store a limited amount of energy over a certain time, if our RES generates more energy than the sum of the energy which home appliances need and the energy which is able to be stored in the ESS in a time slot, the remaining energy of RES will be wasted. Hence, (\ref{RES_elements_1}) should be changed to

\begin{equation}
E_{RES}(t) \geq E_{RES}^{load}(t) + E_{RES}^{charge}(t).
\label{RES_elements_2}
\end{equation}

\subsection{Energy Storage System}
As described in Fig. \ref{electricity_flow}, our ESS is charged by the RES energy and energy from the main grid and is discharged for home load use and selling. Hence, with $\forall t, \hspace{0.1cm}1 \leq t \leq T$, we have the following formulas.

\begin{equation}
E_{ESS}^{Discharge}(t)= E_{ESS}^{load}(t) + E_{ESS}^{selling}(t)
\label{ESS_Discharge}
\end{equation}
\begin{equation}
E_{ESS}^{Charge}(t)= E_{RES}^{charge}(t) + E_{MG}^{charge}(t)
\label{ESS_Charge}
\end{equation}
where $E_{ESS}^{Discharge}(t)$ refers to the energy quantity which is drawn from the ESS in a time slot $t$. $E_{ESS}^{Charge}(t)$ refers to the energy quantity stored in the ESS in a time slot $t$. $E_{ESS}^{load}(t)$ is the energy quantity used for home load in a time slot $t$. $E_{ESS}^{selling}(t)$ is the energy quantity sold to the outside in a time slot $t$. $E_{RES}^{charge}(t)$ is the energy quantity stored in the ESS from the RES in a time slot $t$. $E_{MG}^{charge}(t)$ is the energy quantity stored in the ESS from the main grid in a time slot $t$.

The energy level of the ESS after time slot $t$ becomes
\begin{align}
E_{ESS}^{Level}(t)= E_{ESS}^{Level}(t-1) &+ E_{ESS}^{Charge}(t) \cdot \eta^{ESS} \nonumber \\
&- E_{ESS}^{Discharge}(t) / \eta^{ESS} 
\label{ESS_Level}
\end{align}
where $\eta^{ESS}$ is the ESS efficiency. As shown in (\ref{ESS_Level}), when we charge or discharge ESS, we lose small amount of energy which depends on the parameter $\eta^{ESS}$.

When using the ESS, we must satisfy the following constraints.
\begin{itemize}
\item The charge/discharge rate of the ESS cannot exceed the $Ch_{rate}/Dh_{rate}$. This means that we are only able to put in or draw a certain maximum energy quantity in a time slot $t$ with duration $\Delta t$.
\item The energy level of the ESS must be between $EL_{min}$ and $EL_{max}$.
\item We should avoid simultaneous charging and discharging of our ESS.
\end{itemize}
From the above, we have the following constraints.

\begin{equation}
0 \leq E_{ESS}^{Discharge}(t) \leq Dh_{rate} \cdot \Delta t \cdot \big(1 - mode^{ESS}(t)\big)
\label{ESS_Discharge_Contraints}
\end{equation}
\begin{equation}
0 \leq E_{ESS}^{Charge}(t) \leq Ch_{rate} \cdot \Delta t \cdot mode^{ESS}(t)
\label{ESS_Charge_Contraints}
\end{equation}
\begin{equation}
EL_{min} \leq E_{ESS}^{Level}(t) \leq EL_{max}
\label{ESS_Level_Contraints}
\end{equation}
\begin{equation}
mode^{ESS}(t) = 
   \begin{cases}
   1 & \quad \text{if the ESS is charged in time slot } t \\
   0 & \quad \text{if the ESS is discharged in time slot } t
   \end{cases}
\label{charge_discharge_ESS_mode}
\end{equation}

Combining (\ref{ESS_Discharge}), (\ref{ESS_Charge}), (\ref{ESS_Discharge_Contraints}), and (\ref{ESS_Charge_Contraints}), we have the following constraints for $E_{ESS}^{load}(t)$, $E_{ESS}^{selling}(t)$, $E_{RES}^{charge}(t)$, and $E_{MG}^{charge}(t)$.

\begin{equation}
0 \leq E_{ESS}^{load}(t) + E_{ESS}^{selling}(t) \leq Dh_{rate} \cdot \Delta t \cdot \big(1 - mode^{ESS}(t)\big)
\label{upper_bound_discharge}
\end{equation}
\begin{equation}
0 \leq E_{RES}^{charge}(t) + E_{MG}^{charge}(t) \leq Ch_{rate} \cdot \Delta t \cdot mode^{ESS}(t)
\label{upper_bound_charge}
\end{equation}
\

Because the sum of $E_{ESS}^{load}(t)$ and $E_{ESS}^{selling}(t)$ has an upper bound as shown in (\ref{upper_bound_discharge}), there is a trade-off between $E_{ESS}^{load}(t)$ and $E_{ESS}^{selling}(t)$. When $E_{ESS}^{load}(t)$ increases, $E_{ESS}^{selling}(t)$ will decrease, and vice versa. Likewise, there is a trade-off between $E_{RES}^{charge}(t)$ and $E_{MG}^{charge}(t)$.

Since we only consider our system over the course of day (with no net accumulation being carried over to the next day), the energy level must be returned to the initial energy level by the end of the day. Thus, we have this constraint.

\begin{equation}
E_{ESS}^{Level}(T)= EL_0
\label{ESS_last_level_Constraints}
\end{equation}

We assume that all energy to be sold comes from the ESS. The parameters of our ESS used in this paper are shown in Table \ref{ESS_parameters}.

\begin{table}[h]
\caption{The parameters of an ESS.}
\centering
\begin{tabular}{|c|c|}
\hline
\textbf{Parameters}	& \textbf{Meaning}\\
\hline
$\eta^{ESS}$ & ESS efficiency \\
$Ch_{rate}/Dh_{rate}$ & maximum charge/discharge rate of the ESS \\
$EL_0$ & initial energy level of the ESS \\
$EL_{min}$ & minimum energy level of the ESS \\
$EL_{max}$ & maximum energy level of the ESS \\
\hline
\end{tabular}
\label{ESS_parameters}
\end{table} 

\subsection{Home Appliances}
In our system, we suppose that there are two different sets of appliances: shiftable appliances $M$ and non-shiftable appliances $N$. The set of shiftable devices $M=\{a_1,a_2,a_3,...,a_m\}$ includes the devices which can operate during any time slot whereby we can move the operation time of these devices to low price slots to save cost. The set of non-shiftable devices $N=\{b_1,b_2,b_3,...,b_n\}$ includes the devices which have fixed operation time slots defined by users. None of the appliances can be interrupted during their operation.

The operation time of each non-shiftable appliance $b_i$ is defined by binary parameter $O_{b_i}(t)$ which shows the status of device $b_i$ in time slot $t$. These parameters have a fixed value and are defined by users.

\begin{equation}
O_{b_i}(t) = 
   \begin{cases}
   1 & \quad \text{if non-shiftable device  $b_i$ is ON} \\
   0  & \quad \text{if non-shiftable device $b_i$ is OFF}
   \end{cases}
\end{equation}

Assuming $PR_{b_i}$ is the power rating of device $b_i$ given by producers, the energy consumption of non-shiftable set $N$ in a time slot $t$ is calculated as

\begin{equation}
E_N(t)= \displaystyle\sum_{i=1}^{n}\Big(PR_{b_i} \times O_{b_i}(t) \times \Delta t\Big).
\label{un_shitable_energy_consumption}
\end{equation}

The operation time of each shiftable appliance $a_i$ is defined by one parameter and two variables: parameter $LoT_{a_i}$ is the length of operation time in a day, integer variable $st_{a_i}$ refers to the time slot in which device $a_i$ starts to run, and binary variable $O_{a_i}(t)$ shows the status of device $a_i$ in time slot $t$.

\begin{equation}
O_{a_i}(t) = 
   \begin{cases}
   1 & \quad \text{if shiftable device  $a_i$ is ON} \\
   0  & \quad \text{if shiftable device $a_i$ is OFF}
   \end{cases}
\label{shiftable_appliance_constraint}
\end{equation}

In a time slot $t$, the energy consumption of shiftable set $M$ is calculated as

\begin{equation}
E_M(t)= \displaystyle\sum_{i=1}^{m}\Big(PR_{a_i} \times O_{a_i}(t) \times \Delta t\Big).
\label{shiftable_energy_consumption}
\end{equation}
where $PR_{a_i}$ refers to the power rating of devices $a_i$ given by producers. There are several constraints which these variables must follow. First, each shiftable device must finish its operation within that day.

\begin{equation}
st_{a_i} \leq T - LoT_{a_i} + 1
\label{constraint_starting_time}
\end{equation}

Second, every shiftable device must not be interrupted during its operation time. This means that the binary variable $O_{a_i}(t)$ must satisfy the following constraint.

\begin{equation}
O_{a_i}(t) = 
   \begin{cases}
   1 & \quad \text{if } st_{a_i} \leq t \leq st_{a_i} + LoT_{a_i} - 1 \\
   0  & \quad \text{otherwise}
   \end{cases}
\label{constraint_no_interrupt}
\end{equation}

Third, a shiftable device $a_i$ should be started after the device $a_j$'s operation is completed (consecutive tasks). As an example, the clothes dryer should be started after the washing machine has already stopped. We have the following constraint.

\begin{equation}
st_{a_j} + LoT_{a_j} + D_{ji}\leq st_{a_i}
\label{constraint_consecutive_task}
\end{equation}
where parameter $D_{ji}$ is the minimum delay between the time device $a_j$ is stopped and the time device $a_i$ is started and this parameter is determined by the residents.

In a time slot $t$ ($1 \leq t \leq T$), the energy consumption of all the appliances in the house, $E_{total}^{appliances}(t)$, is the sum of the energy consumption of the shiftable set $M$, $E_M(t)$, and non-shiftable set $N$, $E_N(t)$. We have:

\begin{equation}
E_{total}^{appliances}(t) = E_N(t) + E_M(t).
\label{total_energy_consumption_in_time_slot}
\end{equation}

To provide enough energy for home appliances in time slot $t$, we use three different sources: energy from RES, ESS, and the main grid as shown in Fig. \ref{electricity_flow}. Hence, we have the following formula.  

\begin{equation}
E_{total}^{appliances}(t) = E_{RES}^{load}(t) + E_{ESS}^{load}(t) + E_{MG}^{load}(t) 
\label{total_energy_consumption_in_time_slot_2}
\end{equation}

From (\ref{total_energy_consumption_in_time_slot}), we have

\begin{equation}
E_{MG}^{load}(t) + E_{RES}^{load}(t) + E_{ESS}^{load}(t) = E_N(t) + E_M(t). 
\label{energy_main_load_consumption_2}
\end{equation}

As $E_{MG}^{load}(t) \geq 0$ and we assume that the main grid always provides enough electricity for the requirements of our home load, we have the following constraint.

\begin{equation}
0 \leq E_{RES}^{load}(t) + E_{ESS}^{load}(t) \leq E_N(t) + E_M(t)
\label{home_appliance_constraint}
\end{equation}
\subsection{Objective Functions}
\subsubsection{Objective 1: Minimizing energy cost}
We assume that the energy from the RES and ESS is complimentary, thus, we only need to consider energy from the main grid. Whereby in a given time slot $t$, the load needed from main grid, $E_{LD}(t)$, is calculated as

\begin{equation}
E_{LD}(t) = E_{MG}^{load}(t) + E_{MG}^{charge}(t).
\end{equation}

From (\ref{energy_main_load_consumption_2}), we have

\begin{equation}
E_{LD}(t) = E_N(t) + E_M(t) + E_{MG}^{charge}(t) - E_{RES}^{load}(t) - E_{ESS}^{load}(t).
\label{energy_consumption_a_day2}
\end{equation}

In addition, in the time slot $t$, we sell an amount of energy, $E_{ESS}^{selling}(t)$, to the outside. Hence, the energy cost in the time slot $t$, $EC(t)$, is

\begin{equation}
EC(t) = E_{LD}(t) \times P_{MG}(t) - E_{ESS}^{selling}(t) \times P_{sell}(t).
\label{cost_per_time_slot}
\end{equation}
where $P_{MG}(t)$ is the day-ahead price of the main grid in the time slot $t$. This value is determined by the electrical provider and sent to users through the AMI. $P_{sell}(t)$ is the price of selling energy in time slot $t$. This value is decided by residents.
From (\ref{energy_consumption_a_day2}), we have

\begin{align}
EC(t) = &\Big(E_N(t) + E_M(t) + E_{MG}^{charge}(t) - E_{RES}^{load}(t) \nonumber \\
        &- E_{ESS}^{load}(t)\Big)\times P_{MG}(t) - E_{ESS}^{selling}(t) \times P_{sell}(t).
\label{cost_day2}
\end{align}

Since our objective is to minimize the total energy cost over a day, the objective function is defined as

\begin{align}
&min(C_{day})= min\displaystyle\sum_{t=1}^{T}EC(t) = \nonumber \\
&min\displaystyle\sum_{t=1}^{T}\bigg(\Big(E_N(t) + E_M(t) + E_{MG}^{charge}(t) - E_{RES}^{load}(t) \nonumber \\
&- E_{ESS}^{load}(t)\Big) \times P_{MG}(t)- E_{ESS}^{selling}(t) \times P_{sell}(t) \bigg).
\label{objective_cost_function_1}
\end{align}
where $E_N(t)$ has a fixed value and $E_M(t)$ is calculated by (\ref{shiftable_energy_consumption}). Usually, the price of electricity from the main grid is higher than the selling price. We assume that $P_{sell}(t)=\alpha \times P_{MG}(t)$ with $0< \alpha \leq 1$. Thus, the objective function of our system becomes

\begin{align}
min\displaystyle\sum_{t=1}^{T}\bigg(&E_N(t) + E_M(t)+ E_{MG}^{charge}(t) - E_{RES}^{load}(t)\nonumber \\
&- E_{ESS}^{load}(t)- \alpha \times E_{ESS}^{selling}(t)\bigg)\times P_{MG}(t).
\label{objective_cost_function_2}
\end{align}

It is worth noting that our HEMS prefers using ESS energy for home load to selling ESS energy to the outside in order to obtain more profits if $\alpha < 1$ because the sum of $E_{ESS}^{load}(t)$ and $E_{ESS}^{selling}(t)$ has an upper bound as shown in (\ref{upper_bound_discharge}).

\subsubsection{Objective 2: Maximizing user's convenience}
To be able to measure the user's convenience (UC) when a shiftable device is scheduled to run at a specific time, we introduce two kinds of time range which are set by residents for each shiftable device $a_i$: Utilization time range $UTR_{a_i}=[us_{a_i}, ue_{a_i}]$ is the time slots this device can be run; Best time range $BTR_{a_i}=[bs_{a_i}, be_{a_i}]$ is the time slots which  are best suited for operation of this device. If a shiftable device $a_i$ is run in time slot $t$, $UC_{a_i}(t)$ can be calculated as

\begin{equation}
UC_{a_i}(t) = 
   \begin{cases}
   0 & \quad  t \leq us_{a_i}. \\
   \frac{t - us_{a_i}}{bs_{a_i} - us_{a_i}} &\quad us_{a_i} \leq t \leq bs_{a_i}. \\
   1 & \quad bs_{a_i} \leq t \leq be_{a_i}. \\
   \frac{t - ue_{a_i}}{be_{a_i} - ue_{a_i}} &\quad be_{a_i} \leq t \leq ue_{a_i}. \\
   0  & \quad ue_{a_i} \leq t.
   \end{cases}
\label{user_convenience_function}
\end{equation}

\begin{figure}[t]
\centering
\includegraphics[scale=0.4]{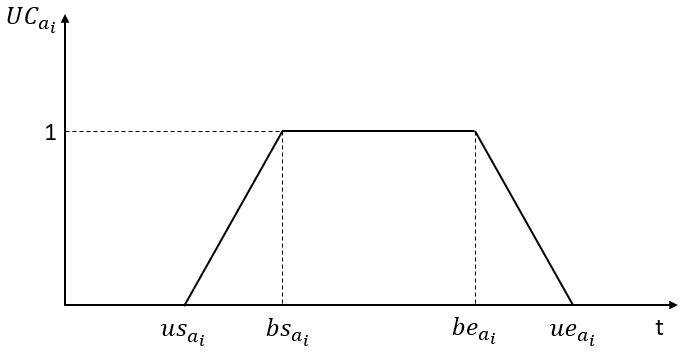}
\caption{Definition of a user's convenience.}
\label{user_convenience_fig}
\end{figure}

Fig. \ref{user_convenience_fig} shows the distribution function $UC_{a_i}(t)$. As shown in Fig. \ref{user_convenience_fig}, when a shiftable device $a_i$ is run in a time slot $t$, the user’s convenience of this device, $UC_{a_i}(t)$, is maximum value of $1$ if the time slot $t$ is in its best time range and is minimum value of $0$ if the time slot $t$ is outside its utilization time range. If the time slot $t$ is inside the utilization time range but outside the best time range, the user's convenience will decrease linearly. Hence, our HEMS tries to maximize user's convenience during a day by scheduling this device inside the best time range or at least the utilization time range.

Our objective function of a user's convenience can be determined as

\begin{equation}
max(UC) = max \displaystyle\sum_{i=1}^{m}\displaystyle\sum_{t=1}^{T} \Big(pri_{a_i}\times UC_{a_i}(t)\times O_{a_i}(t)\Big).
\end{equation}
where $pri_{a_i} \in [1,2,3]$ is the priority of devices $a_i$ that has a highest priority of 3 down to a lowest priority of 1, and this parameter is defined by users.

\subsubsection{Objective 3: Minimizing PAR}
PAR, related to the operation of the main grid, is the ratio of the peak load demand and the average of total load demand over a day. In our system, our objective function of PAR can be defined as

\begin{equation}
min(PAR) = min\Bigg(\frac{(E_{LD}(t))_{max}}{\frac{1}{T}\displaystyle\sum_{t=1}^{T}E_{LD}(t)}\Bigg).
\label{PAR_equation}
\end{equation}
where $E_{LD}(t)$ is calculated by (\ref{energy_consumption_a_day2}).

\subsubsection{Objective 4: Minimizing consecutive waiting time}
For our appliances, some devices $(a_j, a_i)$ have to satisfy the same consecutive constraint as in (\ref{constraint_consecutive_task}). For example, the clothes dryer should be started as soon as possible after the washing machine stops. In this case, the user wants to minimize the waiting time between the time the clothes dryer starts and the time the washing machine stops for maximizing user comfort. However, it is inconvenient if we have to wait a long time to start device $a_i$, even though device $a_j$ has already stopped. In this paper, we propose an objective function to minimize this waiting time (WT) as

\begin{equation}
min (WT) = min \displaystyle\sum \big(st_{a_i} - (st_{a_j} + LoT_{a_j} + D_{ji})\big).
\end{equation}

To minimize the objective of consecutive waiting time and satisfy the consecutive constraint, our HEMS tries to start device $a_i$ as soon as possible, after device $a_j$ is stopped and guarantees a minimum delay of $D_{ji}$.

\subsubsection{Optimization Model}
To optimize our HEMS, all mentioned objectives have to be considered. Hence, a multi-objective function (MO Function) is proposed as our system's optimization model.

\begin{equation}
min(\text{MO Function})= min \Bigg( \frac{C_{day}}{UC - PAR - WT} \Bigg)
\label{multi_objective_function}
\end{equation}

This function must be optimized subject to all previously mentioned constraints of the RES, ESS, and appliances.
\section{A lower bound for energy cost}
In this section, a lower bound for the energy cost $C_{day}$ that is described in (\ref{objective_cost_function_2}) is discovered. Clearly, as $0< \alpha \leq 1$, we have

\begin{align}
C_{day} \geq \displaystyle\sum_{t=1}^{T}&\bigg(E_N(t) + E_M(t)+ E_{MG}^{charge}(t) - E_{RES}^{load}(t) \nonumber \\
&- E_{ESS}^{load}(t)- E_{ESS}^{selling}(t)\bigg)\times P_{MG}(t).
\end{align}
Naming: 

\begin{equation}
E(t) = E_{MG}^{charge}(t) - E_{RES}^{load}(t) - E_{ESS}^{load}(t)- E_{ESS}^{selling}(t).
\label{definition_E}
\end{equation}

We then have

\begin{equation}
C_{day} \geq \displaystyle\sum_{t=1}^{T}\bigg( E_N(t) + E_M(t) + E(t) \bigg) \times P_{MG}(t).
\label{minimum_cost_day_1}
\end{equation}
\begin{align}
\Rightarrow C_{day} \geq \displaystyle\sum_{t=1}^{T}\Big(E_N(t) \times P_{MG}(t)\Big) &+ \displaystyle\sum_{t=1}^{T}\Big(E_M(t) \times P_{MG}(t)\Big) \nonumber \\
&+ \displaystyle\sum_{t=1}^{T}\Big(E(t) \times P_{MG}(t)\Big).
\label{minimum_cost_day_2}
\end{align}

Because $E_N(t)$ and $P_{MG}(t)$ are values we already know, $\displaystyle\sum_{t=1}^{T}\Big(E_N(t) \times P_{MG}(t)\Big)$ has a fixed value. From (\ref{shiftable_energy_consumption}) we have

\begin{align}
&\displaystyle\sum_{t=1}^{T}\Big(E_M(t) \times P_{MG}(t)\Big) \nonumber \\
&=\displaystyle\sum_{i=1}^{m}\displaystyle\sum_{t=st_{a_i}}^{st_{a_i}+LoT_{a_i}-1}\Big(PR_{a_i} \times P_{MG}(t) \times \Delta t \Big) \nonumber \\
&= \displaystyle\sum_{i=1}^{m}\Big(PR_{a_i}\times \Delta t \times \displaystyle\sum_{t=st_{a_i}}^{st_{a_i}+LoT_{a_i}-1}P_{MG}(t)\Big) \nonumber \\
&\geq \displaystyle\sum_{i=1}^{m}\Big(PR_{a_i}\times\Delta t\times M_{MG}^{a_i}\Big).
\label{minimum_consumption_shiftable_devices}
\end{align}

where $M_{MG}^{a_i}$ is the minimum value of $\displaystyle\sum_{t=st_{a_i}}^{st_{a_i}+LoT_{a_i}-1}P_{MG}(t)$ and is easily calculated. Hence, the minimum value of $\displaystyle\sum_{t=1}^{T}\Big(E_M(t) \times P_{MG}(t)\Big)$ is also easily calculated. From (\ref{RES_elements_2}), (\ref{ESS_Discharge}), (\ref{ESS_Charge}) and (\ref{definition_E}), we have

\begin{align}
E(t)\geq E_{ESS}^{Charge}(t) - E_{ESS}^{Discharge}(t) - E_{RES}(t).
\label{minimum_ESS}
\end{align}

Thus, we have

\begin{align}
&\displaystyle\sum_{t=1}^{T}\Big(E(t) \times P_{MG}(t)\Big) \nonumber \\
&\geq \displaystyle\sum_{t=1}^{T}\Big(E_{ESS}^{Charge}(t) - E_{ESS}^{Discharge}(t) \Big)\times P_{MG}(t) \nonumber \\
& \hspace{3cm}-\displaystyle\sum_{t=1}^{T}\Big(E_{RES}(t)\times P_{MG}(t)\Big).
\end{align}
where $\displaystyle\sum_{t=1}^{T}\Big(E_{RES}(t)\times P_{MG}(t)\Big)$ is the cost-benefit we achieve from our PV system in a day, this is easily calculated. Hence, the lower bound of $\displaystyle\sum_{t=1}^{T}\Big(E(t) \times P_{MG}(t)\Big)$ only depends on the expression $\displaystyle\sum_{t=1}^{T}\Big(E_{ESS}^{Charge}(t) - E_{ESS}^{Discharge}(t) \Big)\times P_{MG}(t)$. A nice thing about this expression is that it only involves $E_{ESS}^{Charge}(t)$ and $E_{ESS}^{Discharge}(t)$, an energy quantity which can be stored or drawn from ESS in a time slot $t$. These variables only need to satisfy the constraints in (\ref{ESS_Discharge_Contraints}), (\ref{ESS_Charge_Contraints}), (\ref{ESS_Level_Contraints}), (\ref{charge_discharge_ESS_mode}), (\ref{ESS_last_level_Constraints}). Hence, this expression is a simple MIP problem and the minimum of this expression is easily determined using Matlab or Python. In appendix A, a python program is written to calculate this minimum. Naming $C_{MIP}^{min}$ as this minimum, we have

\begin{align}
\displaystyle\sum_{t=1}^{T}\Big(E(t) \times P_{MG}(t)\Big) \geq C_{MIP}^{min}-\displaystyle\sum_{t=1}^{T}\Big(E_{RES}(t)\times P_{MG}(t)\Big).
\label{minimum_ESS_element}
\end{align}

Combining (\ref{minimum_cost_day_2}), (\ref{minimum_consumption_shiftable_devices}) and (\ref{minimum_ESS_element}), a lower bound of the energy cost $C_{day}$ can be determined as

\begin{align}
C_{day} \geq \displaystyle\sum_{t=1}^{T}\Big(E_N(t) &\times P_{MG}(t)\Big) + \displaystyle\sum_{i=1}^{m}\Big(PR_{a_i}\times \Delta t \times M_{MG}^{a_i} \Big) \nonumber \\
&+C_{MIP}^{min}-\displaystyle\sum_{t=1}^{T}\Big(E_{RES}(t)\times P_{MG}(t)\Big).
\label{lower_bound_cost}
\end{align}

The equal sign = is only valid if the equal signs = of (\ref{minimum_consumption_shiftable_devices}), (\ref{minimum_ESS_element}) are valid and $\alpha = 1$. 

During the installation process, when the resident wants to estimate the economic benefits in the HEMS system, the engineer can apply the python program in Appendix A to calculate $C_{MIP}^{min}$ for the ESS and use (\ref{RES_power}) to estimate the cost-benefit from the RES. By applying these values with the minimum energy cost of shiftable appliances and the energy cost of non-shiftable appliances into (\ref{lower_bound_cost}), we can quickly estimate the economic benefit achieved from the HEMS system.
\section{Simulations and Discussions}
In this section, a house with a set of household appliances, a RES and an ESS is simulated over the course of a day. The parameters for the household appliances are shown in Table \ref{appliances} where the units of PR and LoT are kW and hour, respectively. There are 17 appliances that were divided into two categories: shiftable and non-shiftable. The shiftable appliances are devices whose operating time can be shifted to low price time slots whereas operating times of non-shiftable devices cannot be changed. None of the appliances can be interrupted during operation. We further define three consecutive constraints: the clothes dryer (CD) should be started as soon as possible after the washing machine (WM) is finished ($D_{WM,CD}=0$), the hair dryer (HD) should be started as soon as possible after the electric shower (ES) is finished ($D_{ES,HD}=0$), and the dish washer (DW) should be started 1 hour after the rice cooker (RC) is finished ($D_{RC,DW}=1$).

\begin{table*}[t]
\centering
\caption{Description of the appliances.}
\begin{tabular}{|cccccccc|}
\hline
\textbf{Load Type}	& \textbf{Appliances}	& \textbf{PR}  & \textbf{LoT} & \textbf{Start Time} & \textbf{UTR} &\textbf{BTR} &\textbf{pri} \\
\hline
\multirow{12}{*}{Shiftable} 
&Toaster		 & 0.8	& 1   &-  &1 A.M. - 10 A.M. &6 A.M. - 8 A.M.   &3 \\
&Iron		     & 1.1	& 1   &-  &1 A.M. - 1 P.M. &5 A.M. - 7 A.M.   &2 \\
&Vacuum Cleaner	 & 0.7	& 1   &-  &8 A.M. - 8 P.M.  &9 A.M. - 12 A.M.  &2 \\
&Microwave		 & 0.9	& 1   &-  &8 A.M. - 7 P.M.  &11 A.M. - 2 P.M. &3 \\
&Electric Kettle & 1.0  & 1	  &-  &4 A.M. - 12 A.M. &6 A.M. - 7 A.M.   &3 \\
&Air Conditioner & 1.3  & 10  &-  &5 A.M. - 12 P.M. &9 A.M. - 11 P.M.  &2 \\
&Washing Machine & 1.0  & 2   &-  &7 A.M. - 9 P.M.  &8 A.M. - 2 P.M.   &1 \\
&Clothes Dryer   & 1.8  & 1   &-  &9 A.M. - 9 P.M.  &11 A.M. - 5 P.M.  &1 \\
&Rice Cooker     & 0.6  & 2   &-  &3 P.M. - 9 P.M.  &5 P.M. - 8 P.M.   &3 \\
&Dish Washer     & 1.4  & 2   &-  &4 P.M. - 12 P.M. &8 P.M. - 11 P.M.  &2 \\
&Electric Shower & 2.5  & 1   &-  &5 P.M. - 12 P.M. &8 P.M. - 10 P.M.  &2 \\
&Hair Dryer      & 1.0  & 1   &-  &8 P.M. - 12 P.M. &10 P.M. - 11 P.M. &1 \\
\hline
\multirow{5}{*}{non-shiftable} 
&Personal computers		& 0.2	& 14  & 8 A.M. &- &- &-\\
&Security cameras		& 0.1	& 24  & 0 A.M. &- &- &-\\
&Refrigerator		    & 0.9	& 21  & 2 A.M. &- &- &-\\
&Television		        & 0.2	& 6   & 4 P.M. &- &- &-\\
&Lights		            & 0.1	& 7   & 5 P.M. &- &- &-\\
\hline
\end{tabular}
\label{appliances}
\end{table*}

\begin{table}[!t]
\caption{The input parameters of our ESS in the simulation.}
\centering
\begin{tabular}{|c|c|c|c|c|}
\hline
\text{$\eta^{ESS}$}	& \text{$Ch_{rate}/Dh_{rate}$}	& \text{$EL_0$}  & \text{$EL_{min}$} &\text{$EL_{max}$}\\
\hline
 95\% & 1.0 kW & 0.5 kWh & 0.5 kWh & 10 kWh \\
\hline
\end{tabular}
\label{ESS_parameters_2}
\end{table}

\begin{figure}[!t]
\centering
\includegraphics[scale=0.9]{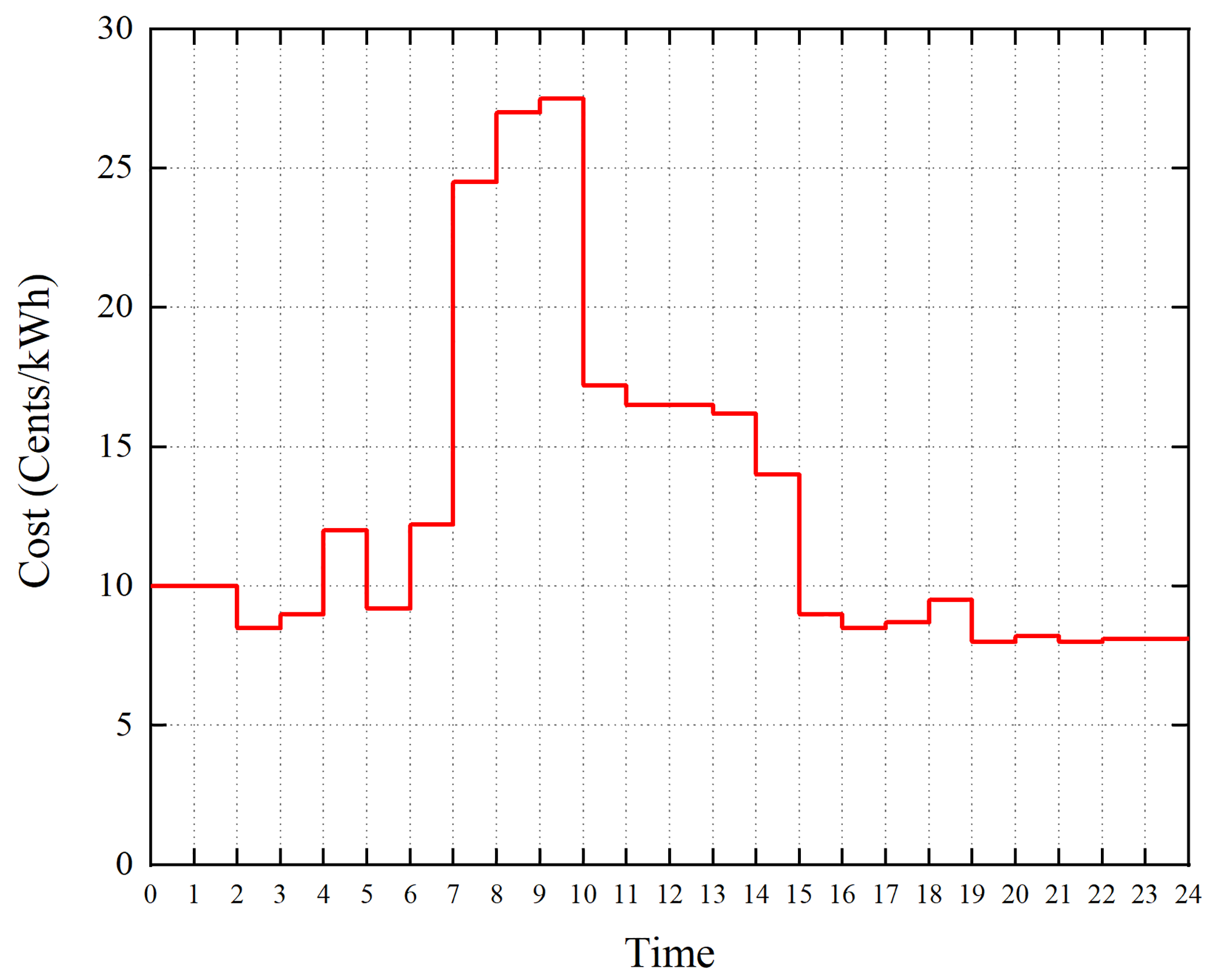}
\caption{Hourly prices according to DAP signal \cite{ahmad2017optimized}.}
\label{DAP_signal}
\end{figure}

\begin{figure}[!t]
\centering
\includegraphics[scale=0.9]{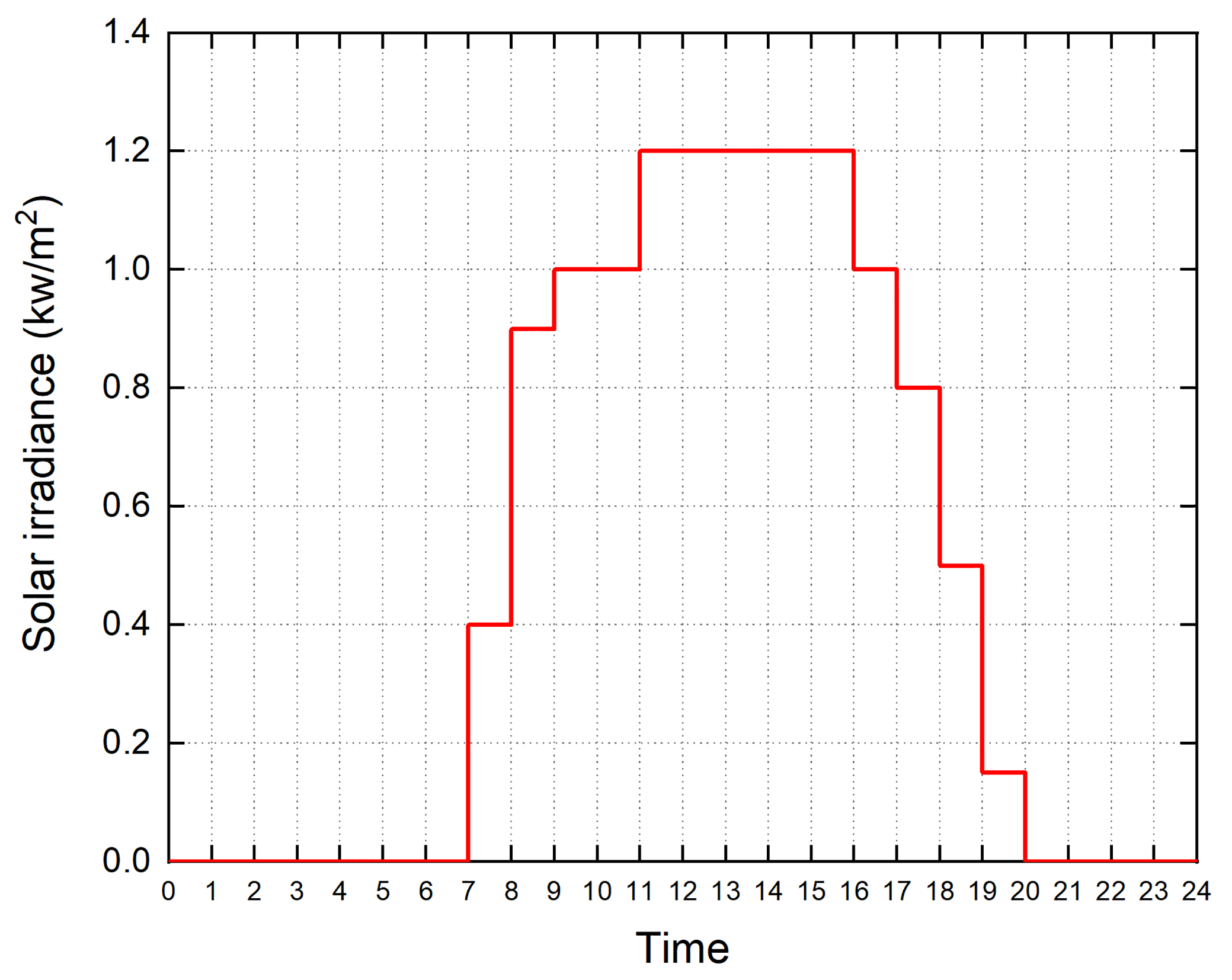}
\caption{Solar irradiance \cite{ahmad2017optimized}.}
\label{Solar_irradiance}
\end{figure}

\begin{figure}[!t]
\centering
\includegraphics[scale=0.9]{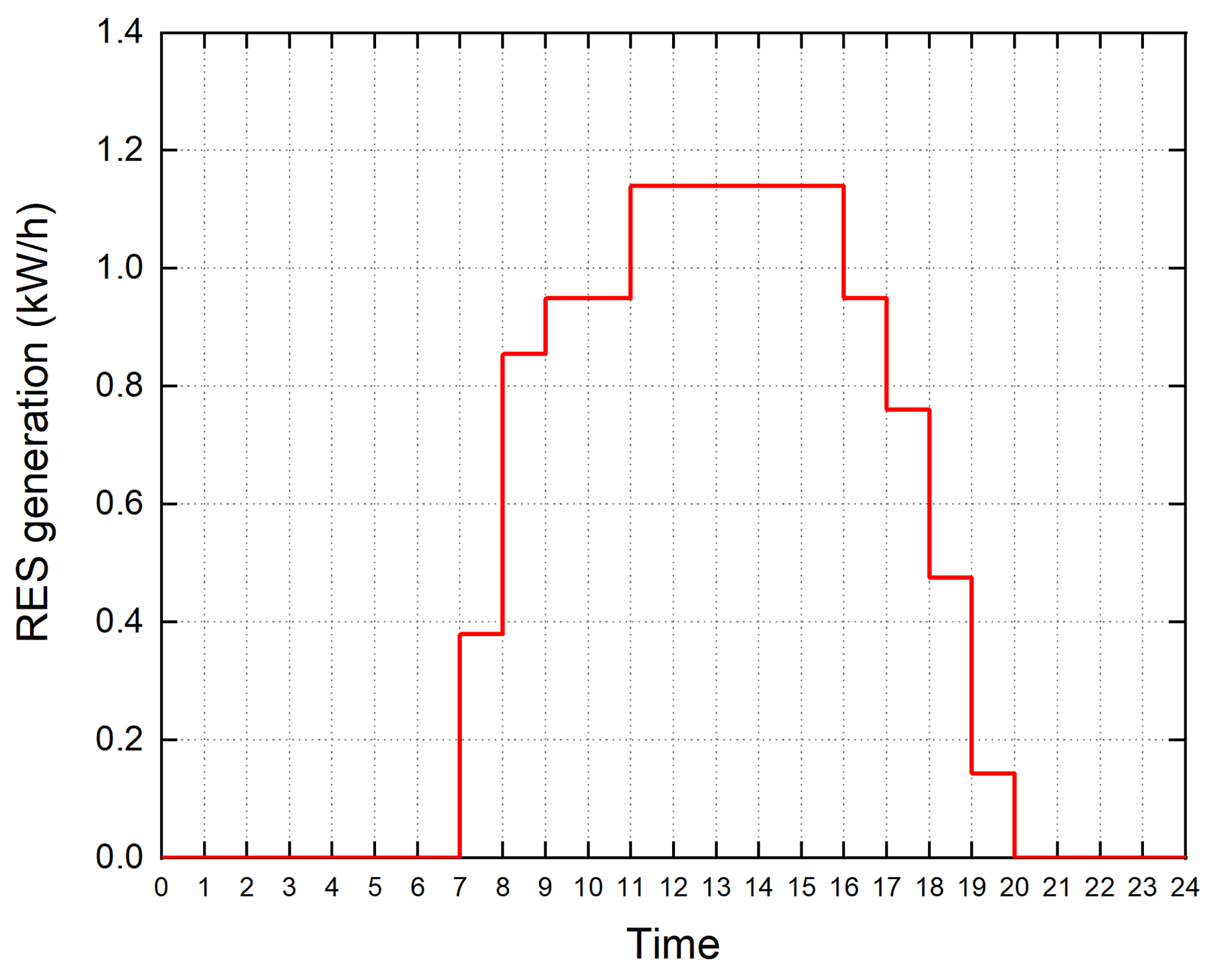}
\caption{Hourly estimated RES.}
\label{RES_generation}
\end{figure}

The parameters for our ESS are shown in Table \ref{ESS_parameters_2}. For the RES in our house, a PV system is used for electricity generation, this is modeled in (\ref{RES_power}). The day ahead price (DAP) and solar irradiance used in our HEMS are shown in Fig. \ref{DAP_signal} and Fig. \ref{Solar_irradiance}. This DAP signal is defined by the electricity provider and the solar irradiance is obtained from METEONORM 6.1 for the Islamabad region of Pakistan \cite{ahmad2017optimized}. We assume that total area of solar panels is $S=1 m^2 $ and the solar conversion efficiency $\eta^{RES}=0.95$. The hourly energy quantity generated by the PV system over the course of a day is shown in Fig. \ref{RES_generation}.

The performance of our HEMS is compared under three different scenarios: $normal, economic,$ and $smart$. The normal scenario describes a situation in which no HEMS is set up. The RES and ESS are also not set up in this scenario. Therefore, there is no ability for utilizing the DAP information, RES, and ESS. Shiftable devices are not controlled according to different objectives and they are run upon the resident's requests. The economic scenario describes a situation in which the HEMS including the RES and ESS is set up. However, shiftable devices are scheduled to minimize energy cost only, other objectives are ignored. On the other hand, in the smart scenario, our HEMS is fully utilized to optimize the multi-objective function, as shown in (\ref{multi_objective_function}). Shiftable devices are scheduled not only to reduce energy cost, but also to optimize the outcomes of other objectives. It is worth noting that in all three scenarios, the consecutive constraints of shiftable devices are always satisfied. All of our simulations were run on an Intel(R) Core(TM) i7-8700 CPU @ 3.20GHz and 16GB RAM with Windows 10 pro (64-bit). The mathematical programming software AIMMS \cite{roelofs2010aimms} with Cplex/Conopt/Outer-Approximation \cite{hunting2011aimms} solvers was used to solve our optimization problem. AIMMS is a high-level modeling system designed for solving LP, NLP, MIP and MINLP problems with minimal user intervention. The computational time taken for the economic and smart scenarios was about 10 seconds and 23 seconds, respectively.
\subsection{Performance of our HEMS in the three scenarios}

\begin{figure}[!t]
\centering
\includegraphics[scale=0.9]{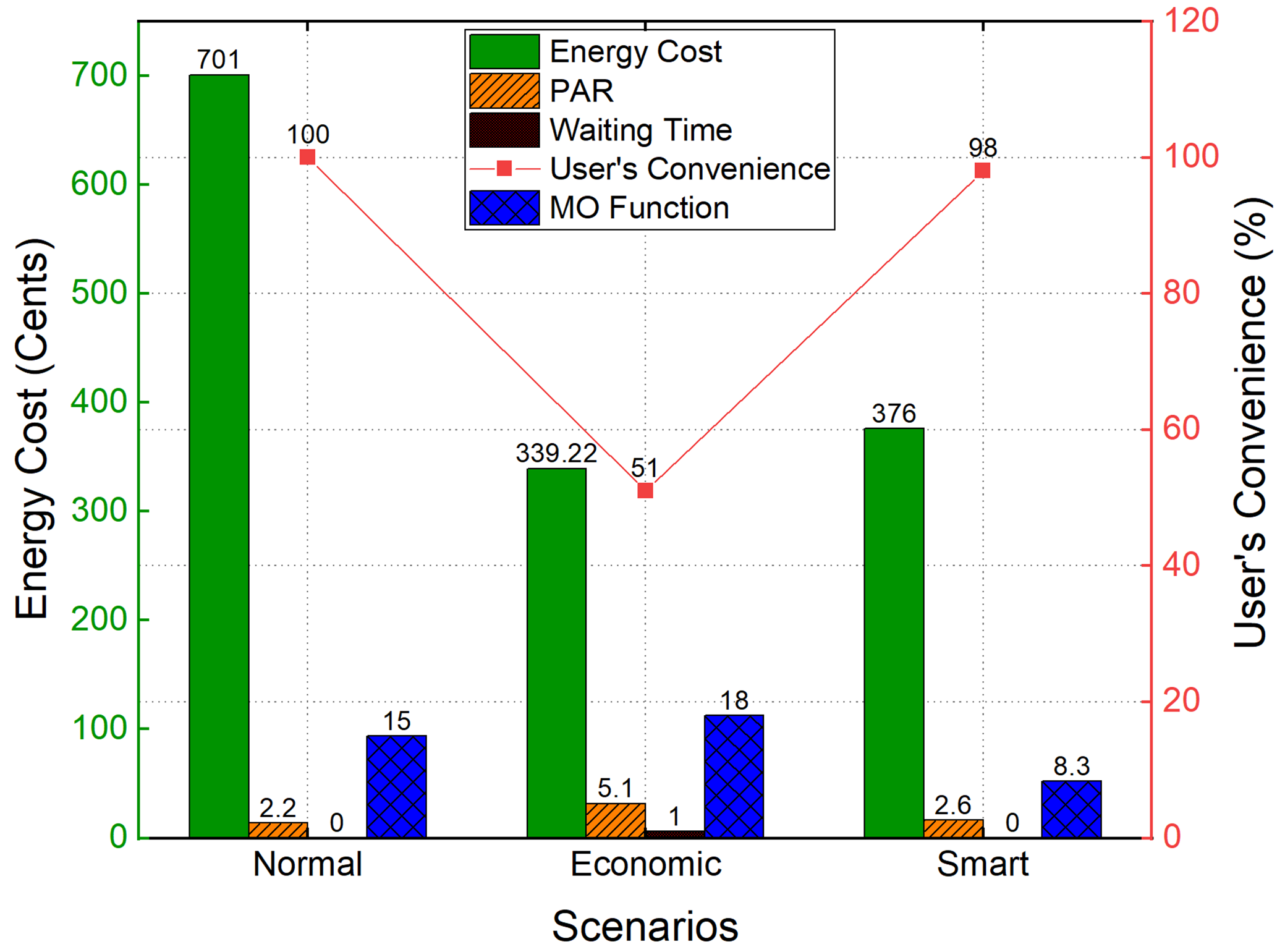}
\caption{Performance of three scenarios with $\alpha = 1$.}
\label{Performance_1}
\end{figure}

Fig. \ref{Performance_1} shows our system's performance for each scenario when $P_{sell}(t) = P_{MG}(t) \hspace{0.1cm}(\forall t\hspace{0.1cm}1 \leq t \leq T)$ which means that $\alpha = 1$. As depicted in this figure, the smart scenario demonstrates excellent performance in comparison with the other scenarios, this is achieved by taking four objectives into account. The multi-objective function value (MO Function) for the smart scenario is only 8.3, a significant improvement of $45\%$ and $54\%$ over the normal and economic scenarios. In detail, in the economic scenario, to achieve the minimum of energy cost (339.22 cents), the user's convenience, consecutive waiting time, and PAR were worst in comparison with the normal scenario. PAR is increased to more than double from $2.2$ to $5.1$ and residents have to wait 1 hour to start to use of shiftable devices. Especially, the economic scenario fails to fulfill the satisfaction of residents in terms of the user's convenience index $(UC/UC_{max}\times 100 )$. The user's convenience in the economic scenario is decreased dramatically from $100\%$ to $51\%$. However, in the smart scenario, with a small increase in energy cost, the user's convenience, consecutive waiting time, and PAR are quite close to those in the normal scenario with a small increase in energy cost.
 
\begin{figure}[!t]
\centering
\includegraphics[scale=0.9]{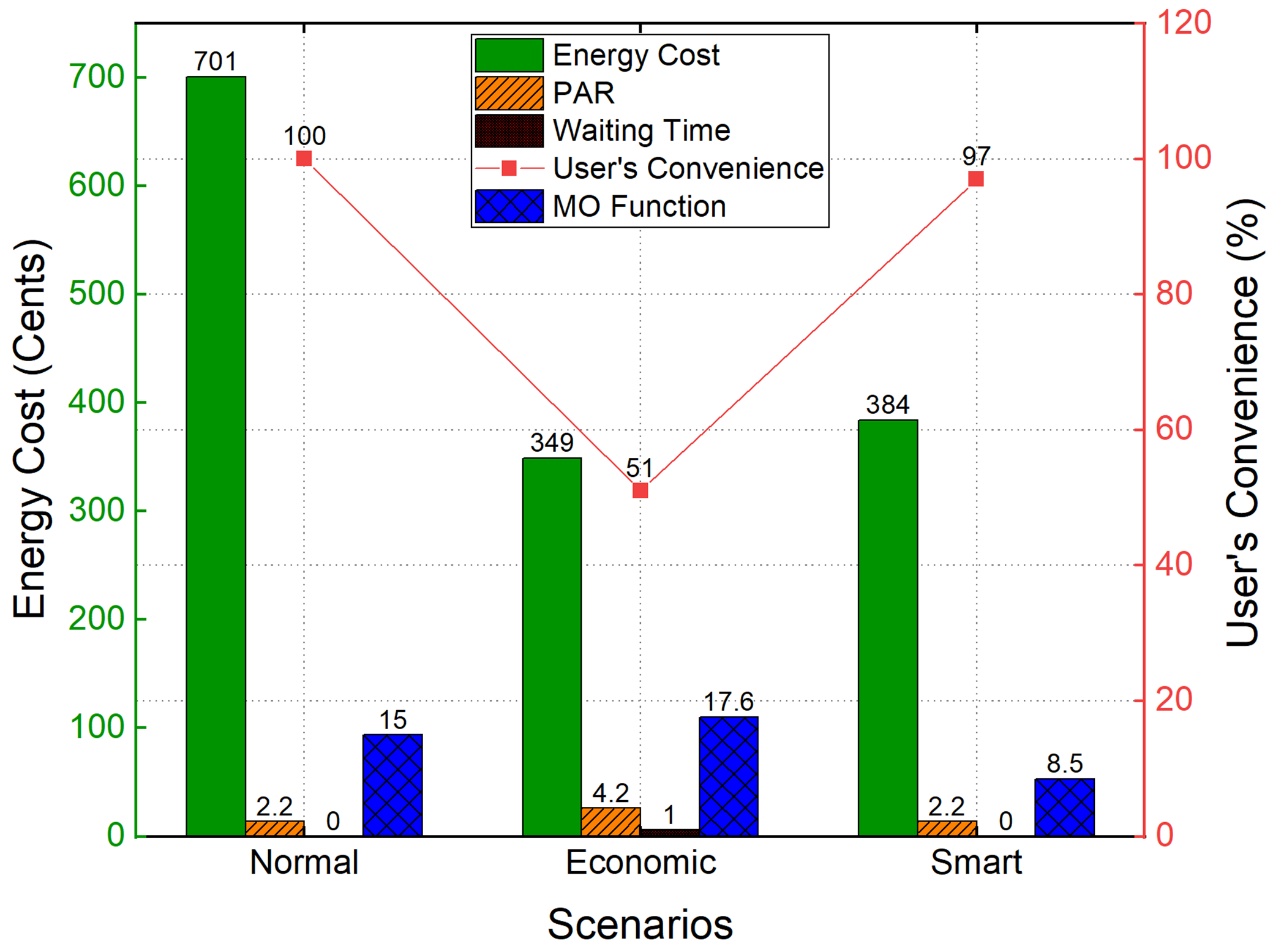}
\caption{Performance of three scenarios with $\alpha = 0.9$.}
\label{Performance_2}
\end{figure}
\begin{figure}[!t]
\centering
\includegraphics[scale=0.9]{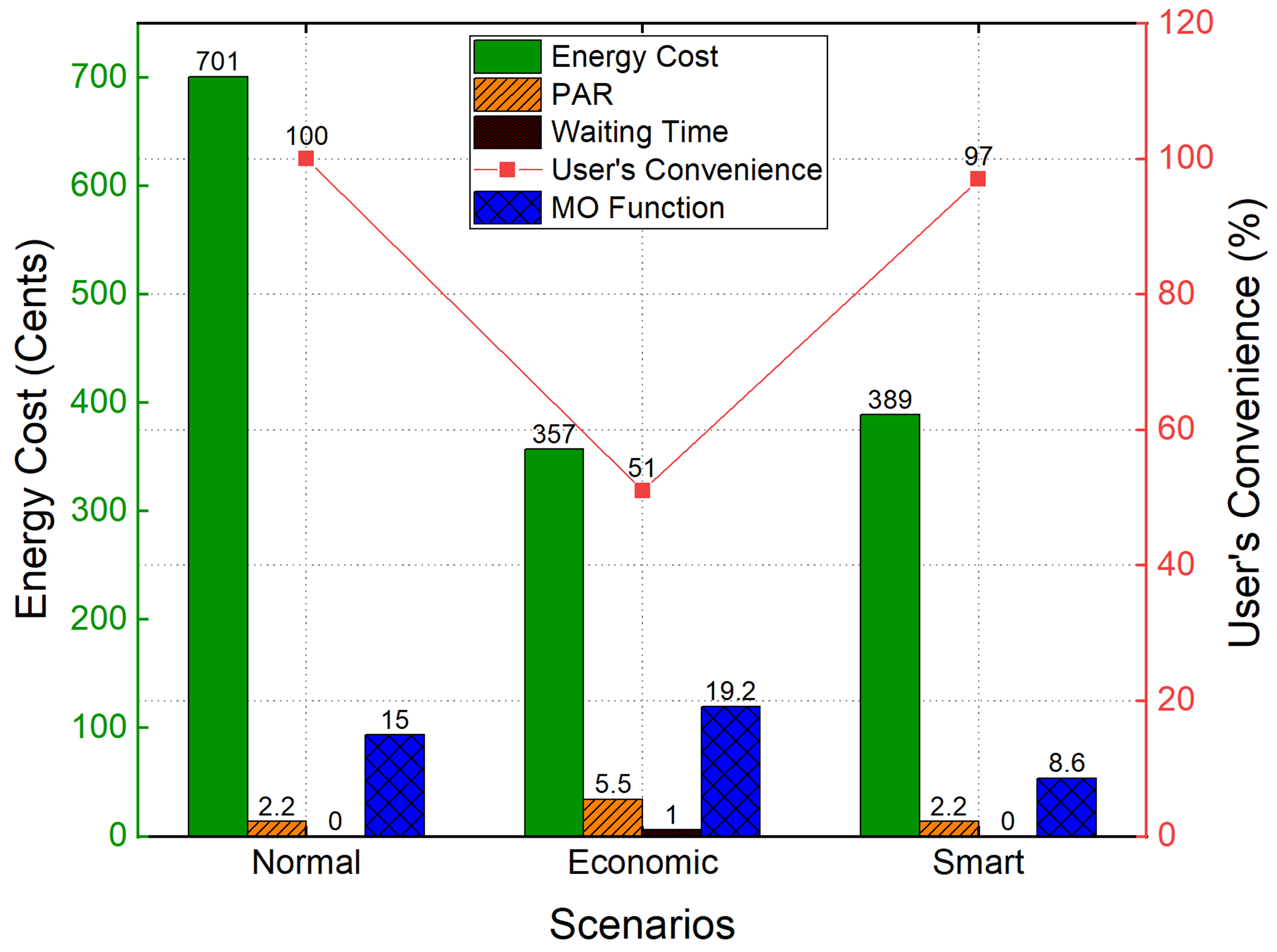}
\caption{Performance of three scenarios with $\alpha = 0.8$.}
\label{Performance_3}
\end{figure}

Moreover, our system still works well in cases where the selling price $P_{sell}(t)$ is decreased as shown in Fig. \ref{Performance_2} and Fig. \ref{Performance_3}. As observed from the simulation results, both the energy cost and multi-objective function value have a tiny increase when $P_{sell}(t)$ is steadily reduced. However, our system still maintains the user's convenience, consecutive waiting time, and PAR like in the normal scenario. In particular, in both cases, the PAR and consecutive waiting time are the same as in the normal scenarios, being $2.2$ and $0$, respectively. Only the user's convenience is slightly decreased to $97\%$. Comparing the smart scenario with $\alpha = 1$ to that of $\alpha = 0.9$ and $\alpha = 0.8$, the reductions of selling price only decrease user's convenience very slightly by $1\%$ and enhance PAR by $0.4$. Moreover, when comparing both smart scenarios with $\alpha = 0.9$ and $\alpha = 0.8$, the user's convenience, consecutive waiting time and PAR, including the scheduling of home appliances, are exactly the same. The only difference between two cases is the increase of energy cost from $384$ cents to $389$ cents.

\begin{figure}[!t]
\centering
\includegraphics[scale=0.9]{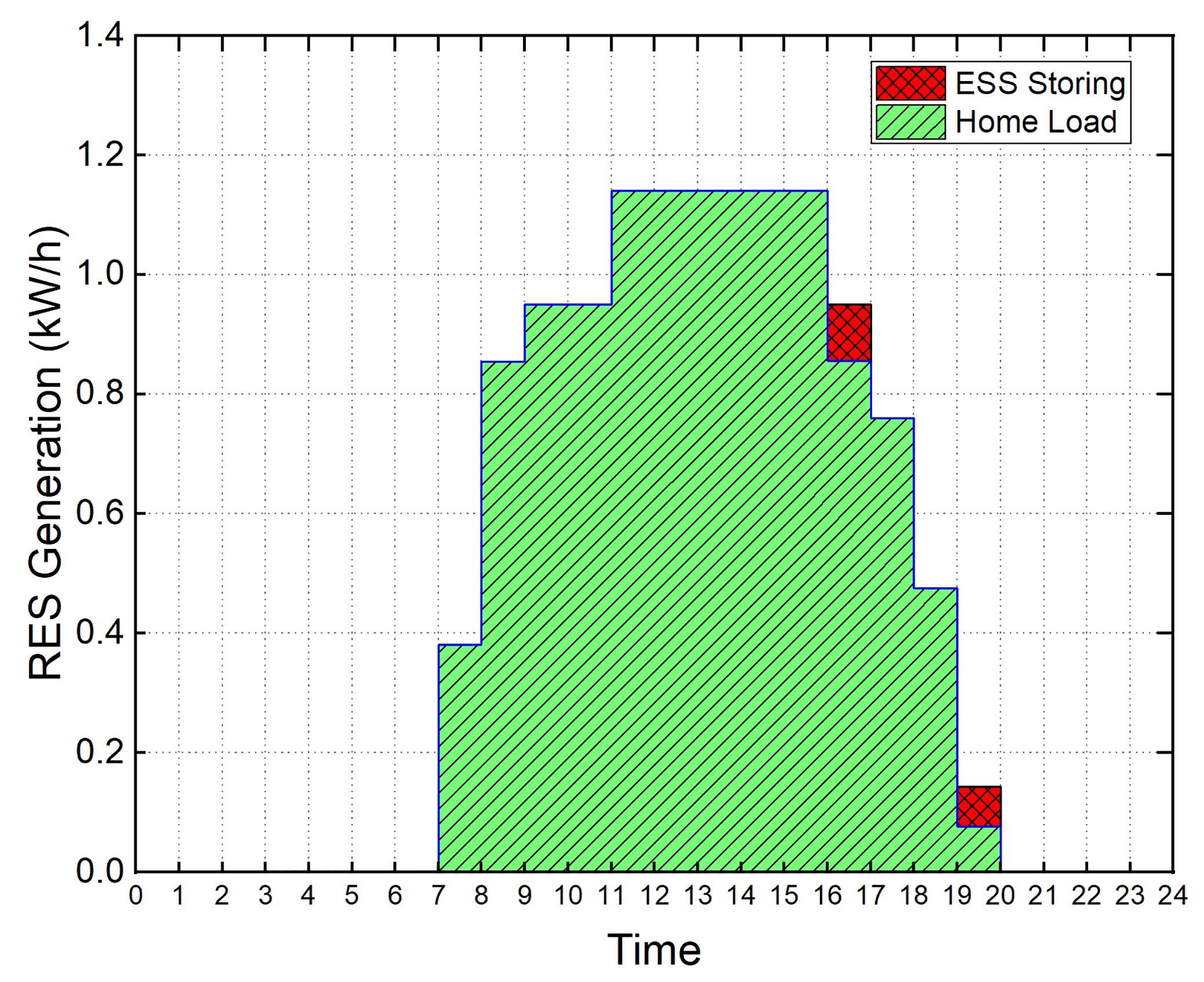}
\caption{Hourly RES usage in smart scenario with $\alpha = 0.9$.}
\label{RES_energy_usage}
\end{figure}

\begin{figure}[!t]
\centering
\includegraphics[scale=0.9]{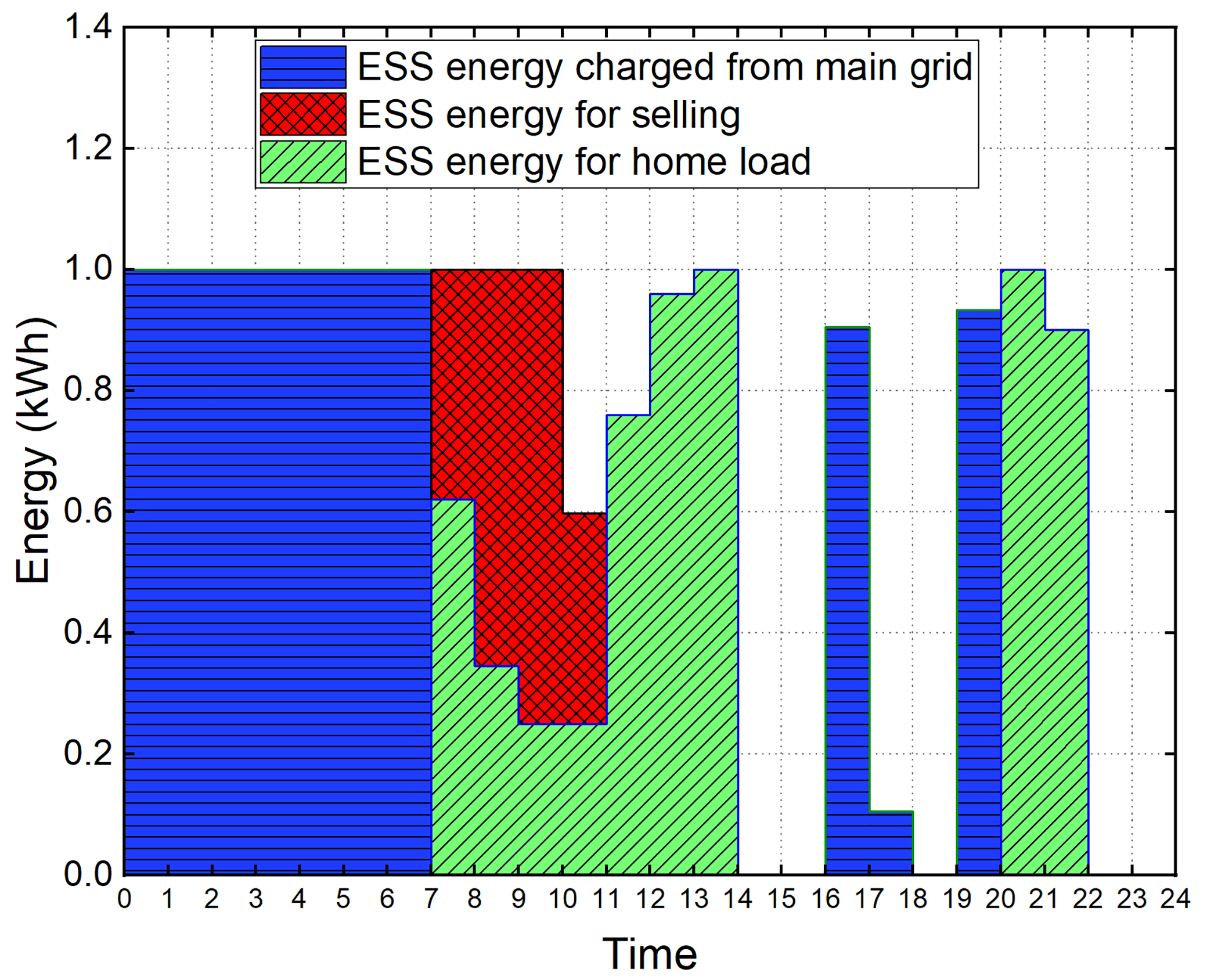}
\caption{Hourly input and output energy of ESS in smart scenario with $\alpha = 0.9$.}
\label{ESS_energy_usage}
\end{figure}

To gain better insights into the performance of the smart scenario, the optimal operation of the RES, ESS, and shiftable appliances in the smart scenario with $\alpha = 0.9$ are shown in Fig. \ref{RES_energy_usage}, Fig. \ref{ESS_energy_usage}, and Table \ref{schedule_shiftable_appliances}. As depicted in Fig. \ref{RES_energy_usage}, because the RES energy is generated during high price time slots, most of it is used for the home load. A small amount of energy is stored in the ESS at time slots from 4 P.M. to 5 P.M. and from 7 P.M. to 8 P.M. In these time slots, the price of the main grid is low and this energy is used for the home load later from 8 P.M. to 10 P.M. By utilizing the RES energy for home load, our HEMS reduces the dependence on the main grid. Moreover, hourly RES energy can be stored in ESS in some time slots and used in next appropriate time slots. This helps our HEMS schedule appliances more flexibly and easily, thereby reducing the energy cost and achieving high user's convenience.

In Fig. \ref{ESS_energy_usage}, in the high price time slots from 7 A.M. to 11 A.M., energy from the ESS is discharged to provide the home load while the surplus energy is also discharged to sell as much as possible to make profits. It is worth noting that, in these time slots, our HEMS only sells ESS energy to the outside after fully supporting home load. This ESS energy came from main grid in the low price time slots from 0 A.M. to 7 A.M. Likewise, the operation of the appliances was scheduled to avoid these high price time slots, as shown in Table \ref{schedule_shiftable_appliances}. Besides, the optimal schedule for the appliances was created by effectively taking the consecutive constraints and time preferences of the residents into account. As an example, the clothes dryer is run after the washing machine finishes. Both of these are operated during the low price times at 3 P.M. and 5 P.M.

\begin{table}[!t]
\caption{Schedule and consumption profile of the shiftable appliances in smart scenario with $\alpha = 0.9$.}
\centering
\begin{tabular}{|cccc|}
\hline
\multirow{2}{*}{\textbf{Appliances}} &\multirow{2}{*}{\textbf{Start Time}} &\textbf{Energy consumption} &\textbf{Cost}   \\

&\textit{} &\textit{(kWh)} &\textit{(cents)} \\
\hline
Toaster			&6 A.M. &0.8 &7.96\\
Iron			    &5 A.M. &1.1 &10.12\\
Vacuum Cleaner		&11 A.M. &0.7 &11.55\\
Microwave		    &12 A.M. &0.9 &14.85\\
Electric Kettle	&6 A.M. &1.0 &12.2\\
Air Conditioner    &1 P.M. &13 &127.66\\
Washing Machine    &3 P.M. &2 &17.5\\
Clothes Dryer      &5 P.M. &1.8 &15.66\\
Rice Cooker		&6 P.M. &1.2 &10.5\\
Dish Washer		&9 P.M. &2.8 &22.54\\
Electric Shower	&8 P.M. &2.5 &20.5\\
Hair Dryer			&9 P.M. &1 &8.0\\
\hline
\end{tabular}
\label{schedule_shiftable_appliances}
\end{table}
\subsection{Lower bound of energy cost}
With energy generated by the RES, as shown in Fig. \ref{RES_generation}, and the set of appliances as shown in Table \ref{appliances}, applying (\ref{lower_bound_cost}), the lower bound of the energy cost in our system was 338.92 cents while the energy cost of economic scenario in our system with $\alpha = 1$ was 339.22 cents, as shown in Fig. \ref{Performance_1}. There is a tiny gap between the two values. To explain where this difference comes from, we consider the schedules of the electrical devices in both scenarios, as shown in Table \ref{plan_comparison_1}.

\begin{table*}[t]
\caption{A comparison of schedules between (\ref{lower_bound_cost}) and economic scenario}
\begin{center}
\begin{tabular}{|c|c|c|c|c|c|c|c|c|}
\hline
\multirow{4}{*}{\textbf{Time Slot}} &\multicolumn{3}{c}{\textbf{Schedule from (\ref{lower_bound_cost})}} &\multicolumn{5}{|c|}{\textbf{Schedule from economic scenario}}\\
\cline{2-9} 
&\textbf{energy demand} &\multicolumn{2}{|c|}{\textbf{ESS}} &\textbf{energy demand} &\multicolumn{4}{|c|}{\textbf{RES, Main Grid, and ESS}}\\
\cline{3-4}
\cline{6-9}
&\textbf{of all appliances} &\text{$E_{ESS}^{Charge}(t)$} &\text{$E_{ESS}^{Discharge}(t)$} &\textbf{of all appliances} &\text{$E_{RES}^{charge}(t)$} &\text{$E_{MG}^{charge}(t)$} &\text{$E_{ESS}^{load}(t)$} &\text{$E_{ESS}^{selling}(t)$}\\
&\textit{(kWh)} &\textit{(kWh)} &\textit{(kWh)} &\textit{(kWh)} &\textit{(kWh)} &\textit{(kWh)} &\textit{(kWh)} &\textit{(kWh)}\\
\hline
0 A.M.-1 A.M.  & 0.1 & 1 & 0 & 0.1 & 0 & 1 & 0 & 0 \\
\hline
1 A.M.-2 A.M.  & 0.1 & 1 & 0 & 0.1 & 0 & 1 & 0 & 0 \\
\hline
2 A.M.-3 A.M.  & 1 & 1 & 0 & 1 & 0 & 1 & 0 & 0 \\
\hline
3 A.M.-4 A.M.  & 1 & 1 & 0 & 1 & 0 & 1 & 0 & 0 \\
\hline
4 A.M.-5 A.M.  & 1 & 1 & 0 & 1 & 0 & 1 & 0 & 0 \\
\hline
5 A.M.-6 A.M.  & 1 & 1 & 0 & 1 & 0 & 1 & 0 & 0 \\
\hline
6 A.M.-7 A.M.  & 1 & 1 & 0 & 1 & 0 & 1 & 0 & 0 \\
\hline
7 A.M.-8 A.M.  & 1 & 0 & 1 & 1 & 0 & 0 & 0.43 & 0.57 \\
\hline
8 A.M.-9 A.M.  & 1.2 & 0 & 1 & 1.2 & 0 & 0 & 0 & 1 \\
\hline
9 A.M.-10 A.M. & 1.2 & 0 & 1 & 1.2 & 0 & 0 & 0 & 1 \\
\hline
10 A.M.-11 A.M. & 1.2 & 0 & 1 & 1.2 & 0 & 0 & 0 & 1 \\
\hline
11 A.M.-12 A.M. & 1.2 & 0 & 1 & 1.2 & 0 & 0 & 0.06 & 0.94 \\
\hline
12 A.M.-1 P.M. & 1.2 & 0 & 1 & 1.2 & 0 & 0 & 0.06 & 0.94 \\
\hline
1 P.M.-2 P.M. & 1.2 & 0 & 0.3175 & 1.2 & 0 & 0 & 0.06 & 0.2575 \\
\hline
2 P.M.-3 P.M. & 2.5 & 0 & 0 & 2.5 & 0 & 0 & 0 & 0 \\
\hline
3 P.M.-4 P.M. & 2.5 & 0 & 0 & 2.5 & 0 & 0 & 0 & 0 \\
\hline
4 P.M.-5 P.M. & 2.7 & 1 & 0 & 2.7 & 0.95 & 0.05 & 0 & 0 \\
\hline
5 P.M.-6 P.M. & 2.8 & 0 & 0 & 2.8 & 0 & 0 & 0 & 0 \\
\hline
6 P.M.-7 P.M. & 2.8 & 0 & 0.9025 & 2.8 & 0 & 0 & 0.9025 & 0 \\
\hline
7 P.M.-8 P.M. & 7.3 & 0 & 0 & 10.6 & 0 & 0 & 0 & 0 \\
\hline
8 P.M.-9 P.M. & 2.8 & 0 & 0 & 4.4 & 0 & 0 & 0 & 0 \\
\hline
9 P.M.-10 P.M. & 11.1 & 0 & 0 & 6.4 & 0 & 0 & 0 & 0 \\
\hline
10 P.M.-11 P.M. & 5.4 & 0 & 0 & 3.8 & 0 & 0 & 0 & 0 \\
\hline
11 P.M.-12 P.M. & 1.5 & 0 & 0 & 2.9 & 0 & 0 & 0 & 0 \\
\hline
\end{tabular}
\label{plan_comparison_1}
\end{center}
\end{table*}

To achieve the lower bound of 338.92 cents for the energy cost calculated by (\ref{lower_bound_cost}), it is required that the input energy $E_{ESS}^{Charge}(t)$ and output energy $E_{ESS}^{Discharge}(t)$ in each time slot have to obey the schedule shown in Table \ref{plan_comparison_1} (to achieve the minimum value of $C_{MIP}^{min}$). As an example, from 0 A.M. to 7 A.M., the ESS has to be in charge mode and $E_{ESS}^{Charge}(t) = 1$ (kWh). Moreover, it is also required that all shiftable appliances have to be run in the lowest price time slots to achieve the minimum value from (\ref{minimum_consumption_shiftable_devices}). The energy demands from both non-shiftable and shiftable appliances in each time slot are shown in this Table. 

\begin{figure}[]
\centering
\includegraphics[scale=0.9]{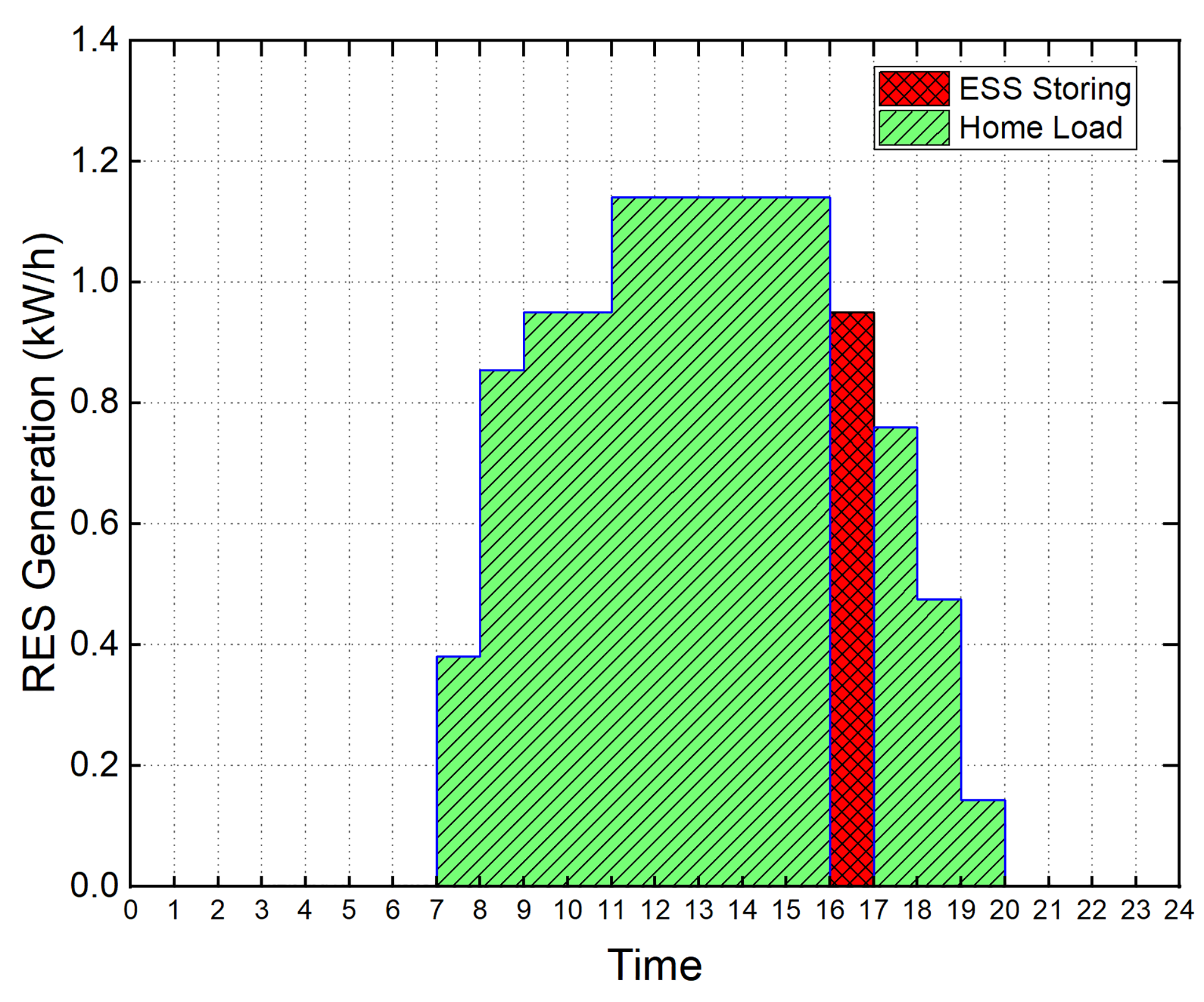}
\caption{Hourly RES usage in economic scenario with $S = 1m^2$}
\label{RES_energy_usage_economic_small}
\end{figure}

In the schedule for the economic scenario, as can be seen in Table \ref{plan_comparison_1}, the input and output energy of the ESS are exactly the same as the schedule from (\ref{lower_bound_cost}). As an example, in the time slot from 7 A.M. to 8 A.M., in schedule from (\ref{lower_bound_cost}), the ESS has to be in discharge mode and $E_{ESS}^{Discharge}(t)= 1$ (kWh). In the economic scenario, the ESS is also in discharge mode and $E_{ESS}^{load}(t) + E_{ESS}^{selling}(t) = 0.43 + 0.57 = 1$ (kWh) in this time slot. There are only small differences in the energy demands of all appliances in the low price time slots from 7 P.M. to 12 P.M. We get these differences because in the economic scenario consecutive constraints are considered while these constraints are not included in (\ref{minimum_consumption_shiftable_devices}). These differences make the energy cost of the economic scenario unequal to the lower bound calculated in (\ref{lower_bound_cost}). It is worth noting that in this economic scenario, during every time slot, the RES energy is always smaller than the energy demand of all appliances and is fully utilized by our system (no loss). The usage of the RES energy is shown in Fig. \ref{RES_energy_usage_economic_small}.

\begin{figure}[]
\centering
\includegraphics[scale=0.9]{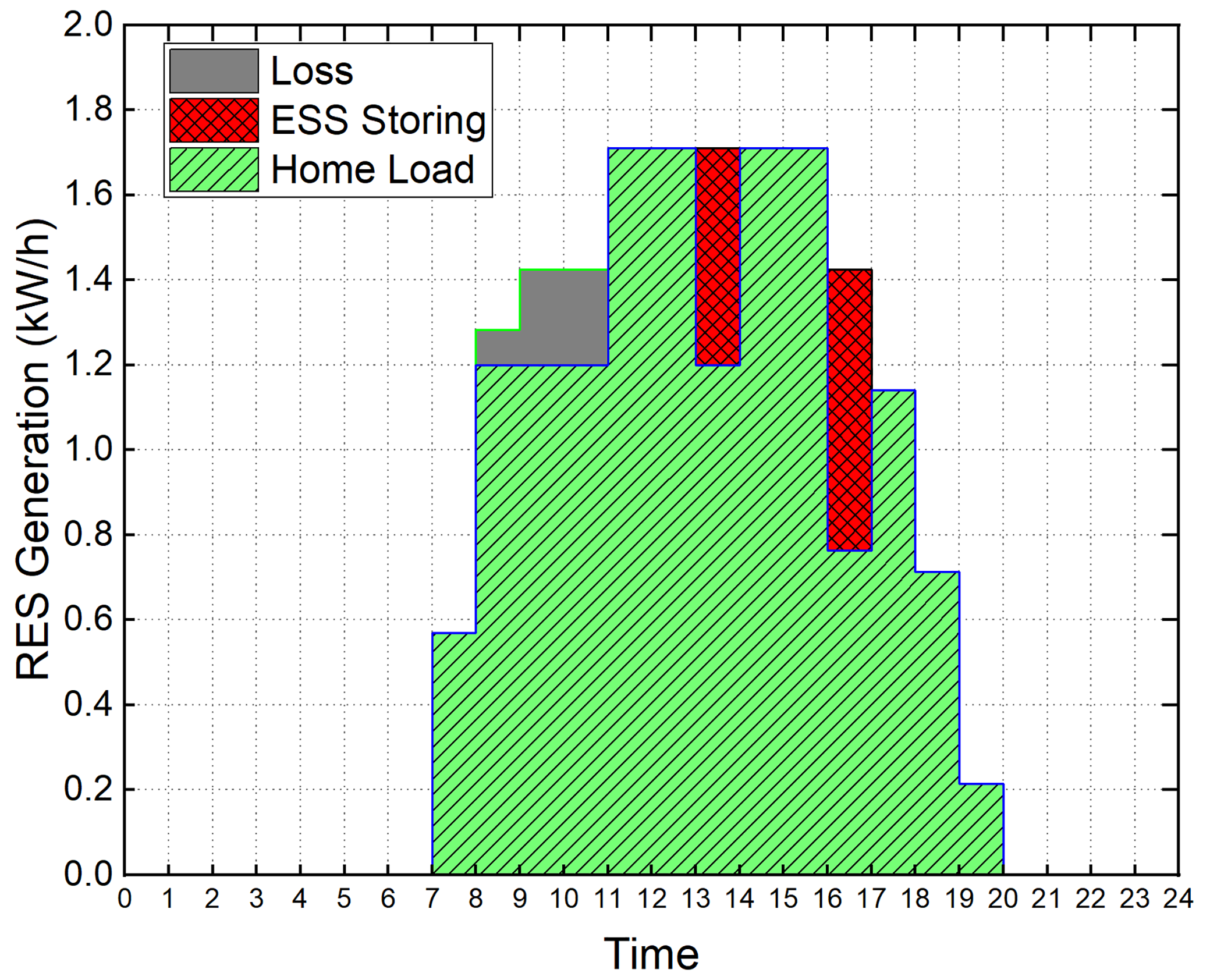}
\caption{Hourly RES usage in economic scenario with $S= 1.5m^2$.}
\label{RES_energy_usage_economic_large}
\end{figure}

Now, we increase the total area of solar panels $S=1.5 m^2$. This means that the energy generated by the RES is increased by $50\%$ in every time slot, as shown in Fig. \ref{RES_energy_usage_economic_large}. In (\ref{lower_bound_cost}), most of the elements are not changed except $E_{RES}(t)$. Thus, the schedule of the appliances and the ESS are not changed and the new lower bound of the energy cost is 250.16 cents. For this case, in the economic scenario, if our ESS and appliances continues following the previous schedule, a lot of RES energy is lost and the energy needed from main grid will be increased dramatically. The reason for this loss is that, from 8 A.M. to 2 P.M, the RES energy is a lot larger than energy demand for all appliances and the ESS can not store the surplus RES energy because the ESS is in discharge mode. Hence, to minimize the energy cost, our system still tries to follow the schedule from (\ref{lower_bound_cost}) but it makes two main changes to reduce the loss of RES energy. First, our system moves some devices to run in the high price time slots from 11 A.M. to 1 P.M. As can be seen in Table \ref{plan_comparison_2}, the energy demand of all appliances in these time slots is increased to 1.8 (kWh). Second, in the time slot from 1 P.M. to 2 P.M., instead of being in discharge mode, our ESS is in charge mode and stores the surplus RES energy (0.51 kWh) as shown in Table \ref{plan_comparison_2}. The main reason for these changes is to consume all the RES energy in these time slots. However, not all RES energy is used. Our system decides to lose some RES energy from 8 A.M. to 11 A.M. as shown in Fig. \ref{RES_energy_usage_economic_large}. Due to the changes and RES loss mentioned above, the new minimum energy cost of our system in this scenario is only 275.2 cents, an increase of $10\%$ compared with the above lower bound of the energy cost.

There is always a gap between lower bound from (\ref{lower_bound_cost}) and the minimum value of the energy cost. If we continue increasing the RES energy by increasing total area of solar panels, this gap will continue to increase and thus the loss of the RES energy will also continue to increase. In real life, residents do not want to lose RES energy, they usually want to set up a PV system which generates an energy quantity equal or a little larger than the energy demand from all appliances for every time slot. As a result, our lower bound is very close to the minimum value of the energy cost and is useful for these houses. It can help residents or engineers estimate quickly how much cost they can save if they want to set up a specific HEMS for a house.

\begin{table*}[t]
\caption{A comparison of schedules between (\ref{lower_bound_cost}) and economic scenario with increasing solar panels.}
\begin{center}
\begin{tabular}{|c|c|c|c|c|c|c|c|c|}
\hline
\multirow{4}{*}{\textbf{Time Slot}}& \multicolumn{3}{c}{\textbf{Schedule from (\ref{lower_bound_cost})}} &\multicolumn{5}{|c|}{\textbf{Schedule from economic scenario}}\\
\cline{2-9} 
& \textbf{energy demand} &\multicolumn{2}{|c|}{\textbf{ESS}} &\textbf{energy demand} &\multicolumn{4}{|c|}{\textbf{RES, Main Grid, and ESS}}\\
\cline{3-4}
\cline{6-9}
&\textbf{of all appliances} &\text{$E_{ESS}^{Charge}(t)$} &\text{$E_{ESS}^{Discharge}(t)$} &\textbf{of all appliances} &\text{$E_{RES}^{charge}(t)$} &\text{$E_{MG}^{charge}(t)$} &\text{$E_{ESS}^{load}(t)$} &\text{$E_{ESS}^{selling}(t)$}\\
&\textit{(kWh)} &\textit{(kWh)} &\textit{(kWh)} &\textit{(kWh)} &\textit{(kWh)} &\textit{(kWh)} &\textit{(kWh)} &\textit{(kWh)}\\
\hline
0 A.M.-1 A.M.  & 0.1 & 1 & 0 & 0.1 & 0 & 1 & 0 & 0 \\
\hline
1 A.M.-2 A.M.  & 0.1 & 1 & 0 & 0.1 & 0 & 1 & 0 & 0 \\
\hline
2 A.M.-3 A.M.  & 1   & 1 & 0 & 1   & 0 & 1 & 0 & 0 \\
\hline
3 A.M.-4 A.M.  & 1   & 1 & 0 & 1   & 0 & 1 & 0 & 0 \\
\hline
4 A.M.-5 A.M.  & 1   & 1 & 0 & 1   & 0 & 1 & 0 & 0 \\
\hline
5 A.M.-6 A.M.  & 1   & 1 & 0 & 1   & 0 & 1 & 0 & 0 \\
\hline
6 A.M.-7 A.M.  & 1   & 1 & 0 & 1   & 0 & 1 & 0 & 0 \\
\hline
7 A.M.-8 A.M.  & 1   & 0 & 1 & 1   & 0 & 0 & 0.43 & 0.57 \\
\hline
8 A.M.-9 A.M.  & 1.2 & 0 & 1 & 1.2 & 0 & 0 & 0 & 1 \\
\hline
9 A.M.-10 A.M. & 1.2 & 0 & 1 & 1.2 & 0 & 0 & 0 & 1 \\
\hline
10 A.M.-11 A.M. & 1.2 & 0 & 1 & 1.2 & 0 & 0 & 0 & 1 \\
\hline
11 A.M.-12 A.M. & 1.2 & 0 & 1 & 1.8 & 0 & 0 & 0.06 & 0.94 \\
\hline
12 A.M.-1 P.M. & 1.2 & 0 & 1 & 1.8 & 0 & 0 & 0.06 & 0.94 \\
\hline
1 P.M.-2 P.M. & 1.2 & 0 & 0.3175 & 1.2 & 0.51 & 0 & 0 & 0 \\
\hline
2 P.M.-3 P.M. & 2.5 & 0 & 0 & 2.5 & 0 & 0 & 0 & 0.7778 \\
\hline
3 P.M.-4 P.M. & 2.5 & 0 & 0 & 2.5 & 0 & 0 & 0 & 0 \\
\hline
4 P.M.-5 P.M. & 2.7 & 1 & 0 & 2.7 & 0.661 & 0.339 & 0 & 0 \\
\hline
5 P.M.-6 P.M. & 2.8 & 0 & 0 & 2.8 & 0 & 0 & 0 & 0 \\
\hline
6 P.M.-7 P.M. & 2.8 & 0 & 0.9025 & 2.8 & 0 & 0 & 0.9025 & 0 \\
\hline
7 P.M.-8 P.M. & 7.3 & 0 & 0 & 8.3 & 0 & 0 & 0 & 0 \\
\hline
8 P.M.-9 P.M. & 2.8 & 0 & 0 & 3.8 & 0 & 0 & 0 & 0 \\
\hline
9 P.M.-10 P.M. & 11.1 & 0 & 0 & 9.5 & 0 & 0 & 0 & 0 \\
\hline
10 P.M.-11 P.M. & 5.4 & 0 & 0 & 3.8 & 0 & 0 & 0 & 0 \\
\hline
11 P.M.-12 P.M. & 1.5 & 0 & 0 & 1.5 & 0 & 0 & 0 & 0 \\
\hline
\end{tabular}
\label{plan_comparison_2}
\end{center}
\end{table*}
\subsection{Weight method used in optimization model}
In (\ref{multi_objective_function}), the user's convenience, consecutive waiting time, and PAR have the same weight coefficient of 1. In Fig. \ref{Performance_1}, in the smart scenario, the PAR of our system is 2.6, we want to reduce this PAR to smaller than the PAR in normal scenario (2.2). To do this job, the weight method of multi-objective optimization (MOO) is used and a new model of optimization is introduced as follows.

\begin{equation}
min(\text{MO Function})= min \Bigg( \frac{C_{day}}{w_1.UC - w_2.PAR - w_3.WT} \Bigg)
\label{multi_objective_function_2}
\end{equation}
where $w_1, w_2, w_3 \in [0,1]$ are the weighting coefficients of user's convenience, PAR, and consecutive waiting time, respectively. These parameters are set by residents and $w_1$ + $w_2$ + $w_3$ = $1$. 
\begin{figure}[]
\centering
\includegraphics[scale=0.9]{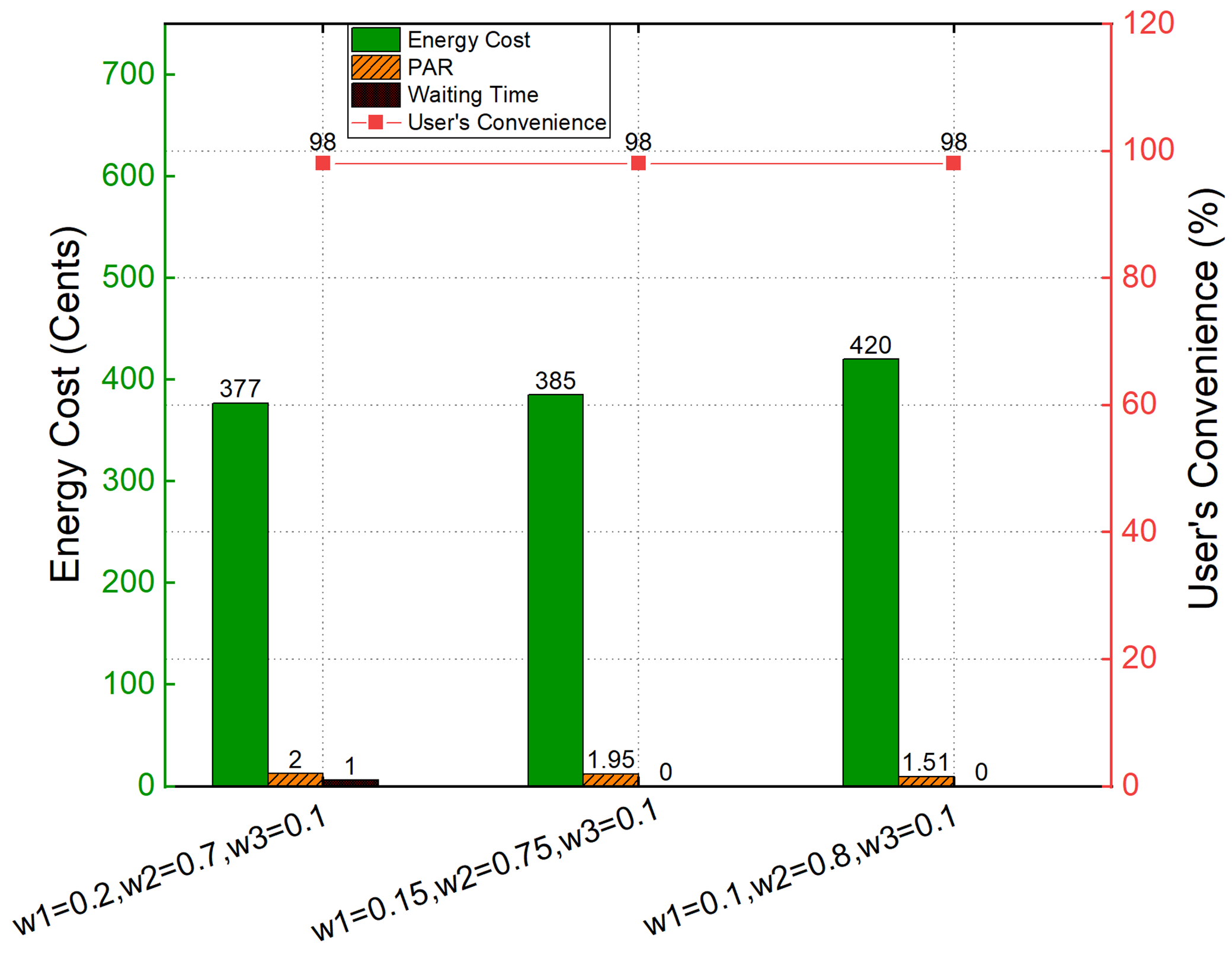}
\caption{The effect of weighting coefficients in our system.}
\label{weight_method}
\end{figure}

As depicted in Fig. \ref{weight_method}, the PAR of our system decreases to 2 when its weight coefficient is set to 0.7, the other weight coefficients for user's convenience and consecutive waiting time are set to 0.2 and 0.1, respectively. If we continue increasing the  weight coefficient of PAR, the value of PAR will continue to decrease. However, there is then a trade-off between the energy cost and PAR. The energy cost of our system will be increased if PAR is decreased. The main reason for this trade-off is that, to decrease PAR, our HEMS has to move operations of some appliances to high price time slots instead of scheduling them during low price time slots. Moreover, as shown in Fig. \ref{weight_method}, even though energy cost of our system is increased, our proposed method keeps the user's convenience stably and changes consecutive waiting time slightly. In other words, the decrease of PAR only affects the energy cost and it does not affect user comfort in our system.

With the weight method of MOO, residents have more flexibility in setting the trade-off between energy cost, user's convenience, consecutive waiting time, and PAR for their homes.

\section{Conclusion}
In this study, a new multi-objective MINLP-based HEMS is mathematically modeled and validated in three different scenarios: normal, economic, and smart. The simulation results show that our HEMS accomplishes a balance among daily energy cost, user's convenience, PAR, and consecutive waiting time. More specifically, the smart scenario shows only slight degradation of user's convenience and PAR by $2\%$ and $18\%$, respectively while achieving $46.4\%$ reduction of daily energy cost and the same level of consecutive waiting time. Furthermore, our simulation results show that a decrease of selling prices has very slight impacts on PAR and user comfort even though the daily energy cost increases. By applying the weight method from MOO, the simulation results show that our system has more flexibility in changing balancing energy cost, user's convenience, consecutive waiting time, and PAR based on the requirements of residents.

In this paper, a lower bound for energy cost was discovered. There is a gap between this lower bound and the minimum value of the energy cost of our system. However, this gap is very small if the energy quantity generated by the RES is equal or a little larger than the energy demands of all the appliances in every time slot. Residents can use this lower bound to quickly calculate an estimate for the cost they can save or choose which parameters for the RES and ESS are suitable for their homes. 

The disadvantage of our proposed HEMS is that it requires residents to configure many parameters of shiftable appliances such as the best time range, the utilization time range, the priority and so on. Moreover, our system cannot reschedule home appliances in order to adapt to the real-time changes which can affect user comfort such as outside temperature or the number of persons in a home. Thus, in the future, we are interested in applying ANN into our HEMS in order to automatically determine the user parameters and adapt to the real-time changes, thereby achieving better performance.

\appendices
\section{Python program to find the minimum value of a MIP problem }
In this appendix, a python program is introduced to find $C_{MIP}^{min}$ in Section IV. Our program uses a library called "mip". Variable $C_{MIP}^{min}$ is found easily by using python software.

\begin{python}
from mip import Model, xsum, BINARY
# price of main grid
price_Grid = [10, 10, 8.5, 9, 12, 9.2, 12.2, 
24.5, 27, 27.5, 17.2, 16.5, 16.5, 16.2, 14, 9, 
8.5, 8.7, 9.5, 8, 8.2, 8, 8.1, 8.1]
# parameters of ESS
effi_ess = 0.95
EL_max = 10
EL_min = 0.5
EL0 = 0.5
number_time_slot = 24
charge_discharge_rate = 1
duration = 24 / number_time_slot

# declaring model
m = Model()

# definition of variables
charge_discharge_mode=[m.add_var(var_type=BINARY)
for i in range(number_time_slot)]
charging_energy = [m.add_var(lb=0) 
for i in range(number_time_slot)]
discharging_energy = [m.add_var(lb=0) 
for i in range(number_time_slot)]
ESS_level = [m.add_var(lb=EL_min, ub=EL_max) 
for i in range(number_time_slot)]

# definition of objective function 
m.objective \
= xsum((charging_energy[i]-discharging_energy[i])
* price_Grid[i] for i in range(number_time_slot))

# adding constraints
for i in range(number_time_slot):
    m += charging_energy[i] <= \
    charge_discharge_rate * duration \
    * charge_discharge_mode[i]
    
    m += discharging_energy[i] <= \
    charge_discharge_rate * duration \
    * (1 - charge_discharge_mode[i])

for i in range(number_time_slot):
    if i > 0:
        m += ESS_level[i] == ESS_level[i - 1] \
        	+ charging_energy[i] * effi_ess \
        	- discharging_energy[i] / effi_ess        
    else:
        m += ESS_level[i] == EL0 \
	        + charging_energy[i] * effi_ess \
    	    - discharging_energy[i] / effi_ess
        
m += ESS_level[23] == EL0
m.optimize()

# print the minimum value
print("optimal value =", m.objective_value)
\end{python}

\bibliographystyle{IEEEtran}
\bibliography{my_references}

\begin{IEEEbiography}[{\includegraphics[width=1in,height=1.25in,clip,keepaspectratio]{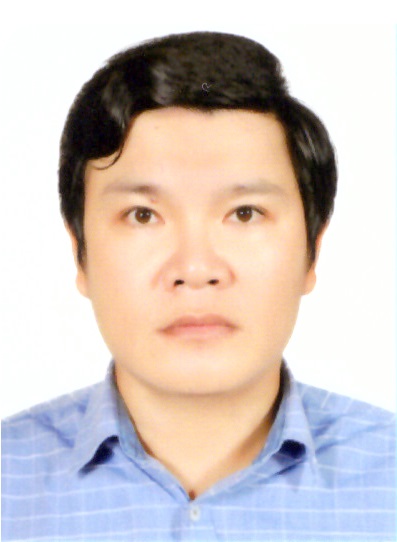}}]{Huy Truong Dinh} was born in Quang Nam, Viet Nam, in 1981. He received the B.S. degree in information technology from Posts and Telecommunications Institute of Technology, Ho Chi Minh, Viet Nam, in 2004
and the M.S. degree in computer engineering from Delft University of Technology, Delft, The Netherlands, in 2009. He is currently pursuing the Ph.D. degree at Soonchunhyang University, Asan, Korea. His research interest includes Internet of Things, energy management, blockchain.
\end{IEEEbiography}
\begin{IEEEbiography}[{\includegraphics[width=1in,height=1.25in,clip,keepaspectratio]{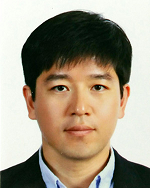}}]{Daehee Kim} received the B.S. degree in electrical and electronic engineering from Yonsei University, Seoul, Korea, in 2003, and the M.S. and Ph.D. degrees in electrical and electronic engineering from Ko-rea University, Seoul, Korea, in 2006 and 2016, respectively. He is currently an Assistant Professor with the Department of Internet of Things, Soonchunhyang University, Asan, Korea. From 2006 to 2016, he was a Senior Engineer with Samsung Electronics, Suwon, Korea, where he conducted research on WiMAX and LTE systems. His research interest includes the Internet of Things, energy management, blockchain, 5G and security for wireless networks.
\end{IEEEbiography}

\EOD

\end{document}